\documentclass[acmsmall]{acmart}

\usepackage{caption}
\usepackage{subcaption}
\usepackage{pifont}%

\usepackage{tabulary}
\usepackage{multirow}
\usepackage{xcolor}
\usepackage{tikz}
\usepackage{enumitem}
\usepackage{amsmath}
\usepackage{tcolorbox}
\usepackage{colortbl}

\newcommand{\notexp}[1]{\textcolor{blue}{XP}}

\newcommand{\MP}[1]{\textcolor{black}{#1}}
\newcommand{\MPB}[1]{\textcolor{blue}{#1}}
\newcommand{\MPRED}[1]{\textcolor{black}{#1}}

\newcommand{\grumbler}[2]{\MPRED{{\bf #1}: #2}}
\newcommand{\newtext}[1]{\MP{#1}}
\newcommand{\renewtext}[1]{\MP{#1}}
\newcommand{\MRtext}[1]{\MP{#1}} %

\newcommand{\chngnum}[1]{}

\AtBeginDocument{%
  \providecommand\BibTeX{{%
    \normalfont B\kern-0.5em{\scshape i\kern-0.25em b}\kern-0.8em\TeX}}}

\setcopyright{rightsretained}
\acmJournal{PACMHCI}
\acmYear{2023} \acmVolume{7} \acmNumber{CSCW2} \acmArticle{254} \acmMonth{10} \acmPrice{}\acmDOI{10.1145/3610045}

\makeatletter
\gdef\@copyrightpermission{
  \begin{minipage}{0.2\columnwidth}
   \href{https://creativecommons.org/licenses/by-nc-nd/4.0/}{\includegraphics[width=0.90\textwidth]{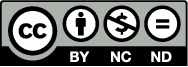}}
  \end{minipage}\hfill
  \begin{minipage}{0.8\columnwidth}
   \href{https://creativecommons.org/licenses/by-nc-nd/4.0/}{This work is licensed under a Creative Commons Attribution-NonCommercial-NoDerivs International 4.0 License.}
  \end{minipage}
  \vspace{5pt}
}
\makeatother

\begin{document}

\title[Cultural Disclosure Norms]{A Tale of Two Cultures: Comparing Interpersonal Information Disclosure Norms on Twitter}

\author{Mainack Mondal }
\affiliation{%
  \institution{IIT Kharagpur}
  \city{Kharagpur}
  \country{India}
}

\author{Anju Punuru}
\affiliation{%
  \institution{IIT Kharagpur}
  \city{Kharagpur}
  \country{India}
}

\author{Tyng-Wen Scott Cheng}
\affiliation{%
  \institution{Brigham Young University}
  \city{Provo}
  \country{US}
}

\author{Kenneth Vargas}
\affiliation{%
  \institution{Brigham Young University}
  \city{Provo}
  \country{US}
}

\author{Chaz Gundry}
\affiliation{%
  \institution{Brigham Young University}
  \city{Provo}
  \country{US}
}

\author{Nathan S Driggs}
\affiliation{%
  \institution{Brigham Young University}
  \city{Provo}
  \country{US}
}

\author{Noah Schill}
\affiliation{%
  \institution{Brigham Young University}
  \city{Provo}
  \country{US}
}

\author{Nathaniel Carlson}
\affiliation{%
  \institution{Brigham Young University}
  \city{Provo}
  \country{US}
}

\author{Josh Bedwell}
\affiliation{%
  \institution{Brigham Young University}
  \city{Provo}
  \country{US}
}

\author{Jaden Q Lorenc}
\affiliation{%
  \institution{Brigham Young University}
  \city{Provo}
  \country{US}
}

\author{Isha Ghosh}
\affiliation{%
  \institution{University of Utah}
  \city{Salt Lake City}
  \country{US}
}

\author{Yao Li}
\affiliation{%
  \institution{University of Central Florida}
  \city{Orlando}
  \country{US}
}

\author{Nancy Fulda}
\affiliation{%
  \institution{Brigham Young University}
  \city{Provo}
  \country{US}
}

\author{Xinru Page}
\affiliation{%
  \institution{Brigham Young University}
  \city{Provo}
  \country{US}
}

\renewcommand{\shortauthors}{Mainack Mondal et al.}

\begin{abstract}

\renewtext{We present an exploration of cultural norms surrounding online disclosure of information about one's interpersonal relationships (such as information about family members, colleagues, friends, or lovers) on Twitter.} 
\renewtext{ The literature identifies the cultural dimension of individualism versus collectivism as being a major determinant of offline communication differences in terms of emotion, topic, and content disclosed. We decided to study whether such differences also occur online in context of Twitter when comparing tweets posted in an individualistic (U.S.) versus a collectivist (India) society. We collected more than 2 million tweets  posted in the U.S. and India over a 3 month period which contain interpersonal relationship keywords. A card-sort study was used to develop this culturally-sensitive saturated taxonomy of keywords that represent interpersonal relationships (e.g., ma, mom, mother).} Then we developed a high-accuracy interpersonal disclosure detector based on dependency-parsing (F1-score: 86\%) \renewtext{to identify when the words refer to a personal relationship of the poster (e.g., "my mom" as opposed to "a mom"). This allowed us to identify the 400K+ tweets in our data set which actually disclose information about the poster's interpersonal relationships.} \newtext{We used a mixed methods approach to analyze these tweets (e.g., comparing the amount of joy expressed about} \renewtext{one's family) and found differences in emotion, topic, and content disclosed between tweets from the U.S. versus India. }
\renewtext{Our analysis also reveals how a combination of qualitative and quantitative methods are needed to uncover these differences; Using just one or the other can be misleading. This study extends the prior literature on Multi-Party Privacy and provides guidance for researchers and designers of culturally-sensitive systems}.

\end{abstract}

\ccsdesc[500]{Security and privacy~Social aspects of security and privacy}
\ccsdesc[300]{Security and privacy~Usability in security and privacy}
\ccsdesc[300]{Security and privacy~Privacy protections}

\keywords{cultural privacy, disclosure norms, contextual integrity, hofstede's dimensions, multi-party privacy, social media privacy}

\received{July 2022}
\received[revised]{January 2023}
\received[accepted]{March 2023}

\maketitle

\section{Introduction}\label{sec:intro}

\noindent Today, social media platforms such as Facebook and Twitter support a global society of users coming from a variety of cultural norms. These platforms mediate information disclosures and interpersonal interactions which play a key role in the formation and co-construction of interpersonal relationships~\cite{petronio_boundaries_2002}. Yet, how one communicates and what is appropriate to share is shaped by social expectations tied to one's culture. Indeed, the literature shows how the social norms around appropriate information disclosure can vary greatly between cultures~\cite{solove-2008-privacyBook}. Cultural and social norms also influence content and style of communication~\cite{samovar_understanding_1981}. With the increasingly global reach of social media, mismatched communication styles and disclosure norms can lead to misunderstanding, hindering the establishment of new relationships, or worse, leading to unintentional conflict. Prior literature already points to how misinterpretation of posts on social media is a source of friction and can even lead to physical danger~\cite{abokhodair-2017-photosharing}. When the poster and reader's information disclosure expectations are shaped by different cultural backgrounds, this could cause further misunderstanding. 

Moreover, while researchers often focus on what people share about themselves, recent work points out how privacy violations are especially problematic when users disclose information about someone else. Indeed, the Multi-Party Privacy (MPP) literature focuses on how rules are negotiated around disclosing personal information that is known by others such as friends and family ~\cite{acquisti2006imagined, acquisti2009privacy, ayalon2013retrospective, bhagat-www10, braunstein2011indirect, exposurecontrol, fang-www10, mondal2017longitudinal, mondal14}. We use the term \textit{interpersonal information disclosure} to refer to disclosing information about someone else. While researchers have made initial strides towards uncovering interpersonal information disclosure norms for U.S or European participants ~\cite{fogues_sharing_2017, such_photo_2017,lampinen_were_2011, such2018multiparty}, little work has explored social norms beyond Western societies. Thus, our study takes a first step towards doing so by comparing interpersonal information disclosure norms between an Asian country, India, and a Western country, the U.S. We focus on information disclosures made on the platform Twitter which is widely used in both countries. Specifically, our \MRtext{overarching research question is: \emph{How do interpersonal information disclosure norms differ between Indian and U.S. tweets?}}

 \MRtext{Drawing on the broader sociological literature, we found that cultural differences in disclosure norms most often trace back to collectivist versus individualist characteristics ~\cite{cho_qualitative_2013,hsu_cross-cultural_2007,schug_relational_2010, khare-2011-brand-india, RAMAMOORTHY2007187}. To explore whether such differences also apply online on the social media platform Twitter, we further analyzed this literature to identify which facets of interpersonal disclosure differ by culture (see Related Works). This led us to focus our research questions on how the content, frequency, and emotion of these disclosures differ. Specifically, we investigate the following research questions in this study:}
\MRtext{
\begin{description}
\item[$RQ1$] How does the frequency of disclosure about different interpersonal relationships (family, friends, co-workers, etc.) on
Twitter differ between India and the U.S?
\item[$RQ2$] How do the topics of interpersonal information disclosures on Twitter differ between India and U.S.?
\item[$RQ3$] How does potentially sensitive information, such as location and financial information, in interpersonal information disclosure on Twitter differ between India and U.S.?
\item[$RQ4$] In the context of interpersonal disclosure, how does the frequency of emotions disclosed in tweets differ between
India and the U.S.?
\item[$RQ5$]  In the context of interpersonal disclosure, how does the frequency of positive/negative emotions disclosed in tweets
differ between India and the U.S.?
\end{description}
}

We took a mixed methods approach to both deductively and inductively investigate these research questions. Our focus was on interpersonal information disclosures on Twitter by posters from the U.S. and India. We collected tweets posted publicly in the U.S. or India over the span of three months and which contained \MRtext{keywords} related to interpersonal information disclosures, resulting in 2,368,547 tweets. \MRtext{To analyze the data, we utilized }\newtext{ various quantitative methods (dependency parsing, Latent Dirichlet Allocation, statistical word counts, deep learning) and qualitative methods (thematic analysis)}. %

 \MRtext{In the process of exploring these research questions, we make contributions to the literature not only in regards to better understanding the answers to these research questions, but also in developing taxonomies and tools that can be used for future work investigating cultural differences in disclosures. More concretely, this work makes the following five contributions to research: }

\textit{Contribution 1: A culturally-sensitive taxonomy of interpersonal relationship words.}
We conducted a theoretically-grounded online card sorting study with 63 U.S. and 58 Indian participants \newtext{who helped us create a saturated list of 178 keywords representing interpersonal relationships (meaning participants were able to add words and relationship categories in the taxonomy until they felt these lists were complete). These were categorized by participants} into 9 broad relationship categories \newtext{(e.g., family, extended family, superiors) which we used to analyze the data.} 
\chngnum{13, 16}
\newtext{Furthermore, we provide a detailed analysis of words that were placed in different relationship categories by participants in the two countries which highlights different perceptions of certain relationships. For example, Indian participants were more likely to place `spouse' within the Family category while U.S. participants placed them in a Lover category. We also observed the inclusion of adoptive and step relationships within the Family category by U.S. participants, as opposed to being considered Extended Family by Indian participants. Although an individualist-collectivist dichotomy might be too broad to capture all subtle nuances of cultural disclosure, our exploration shows that our real-world data does point to differences \MRtext{in relationship taxonomy} due to this high-level dichotomy.}

\textit{Contribution 2: Developing and validating a high-accuracy interpersonal information disclosure detector for large-scale tweet data.}
Second, we leveraged our empirically-derived saturated taxonomy to collect 2,368,547 tweets posted about interpersonal relationships (e.g., tweets containing keywords such as mom, ma, mother) from the U.S. or India. To obtain higher accuracy in identifying tweets disclosing information about one's interpersonal relationships (e.g., "my mom" as opposed to merely mentioning relationship-related keywords such as "a mom"), we created a dependency-parsing based \textit{interpersonal information disclosure detector} to further refine the pool of collected tweets, resulting in over 400k tweets. Our manual validation established that the detector has a high accuracy (F1 score 0.86). \MRtext{Using this dataset, we observed that Twitter users in both cultures prefer to tweet about Family and Friend relationships over more
distant relationships such as Acquaintance, Co-worker, Extended Family, and Supervisor. U.S. tweeters place greater relative emphasis on the Lover, Family
and Extended Family categories, while Indian tweeters place greater relative emphasis on Friend, Best Friend, and
Co-worker relationships.}

\textit{Contribution 3: Identifying culture-specific topics of interpersonal disclosure via a mixed-method analysis.} Third, using a mixed method approach (Latent Dirichlet Allocation-based topic detection~\cite{blei-2003-lda} paired with thematic analysis) we identify the topics that \newtext{were present in interpersonal disclosures for each country}. Our analysis identified 12 topics for U.S. users and 9 topics for Indian users that were frequently the topic of disclosures about interpersonal relationships (ranging from sharing `Stories about Family' to `patriotism'). Some topics were thematically similar across cultures (e.g., schooling, stories about their families, holidays/celebrations), while others show distinctly different disclosure norms, many \MRtext{of which align with} collectivist-individualist dichotomy.

\textit{Contribution 4: Identifying differences in emotional disclosure norms across cultures.} Fourth, we used a deep-learning driven emotion classifier to identify norms of emotional expression in the more than 400,000 tweets identified through our dependency-parsing based interpersonal disclosure detector. Results showed that in the context of interpersonal relationships, Indian tweets contain more joy than U.S. tweets. They also revealed which relationships are spoken about in more or less emotional ways. Our analysis highlights interesting analogous patterns of emotional expression across different relationship groupings within each culture.

\textit{Contribution 5: Examining interpersonal disclosure patterns surrounding potentially privacy-sensitive information.} Finally, \MRtext{guided by prior information disclosure research, we recognized that financial and location information can both be sensitive in western contexts~\cite{cushman_communication_1985}. They also have been found to be sensitive in asian cultures~\cite{chen_differences_1995, tsai2010location, wsjlocation}. Thus, we examined tweets containing financial and location information to identify disclosure norms in the two cultures, to compare and contrast the extent to which people are comfortable sharing this type of information.} \newtext{We found that disclosure patterns varied dramatically when tweets were filtered to consider only disclosures about \textit{someone else's} sensitive data, as opposed to disclosures about sensitive data in general.}\chngnum{12} Strikingly, we also discovered that 30.6\% of location-related Indian tweets involved memories of some form, as contrasted with only 6.6\% of U.S. location tweets.

Taken together, our findings reveal substantial differences in the information disclosure norms \newtext{both \textit{across} cultures and \textit{within} cultures as the relationship context varies. This work informs the ongoing discussions surrounding multiparty privacy and the importance of not only societal, but also interpersonal context in determining the topics, emotional content, and sensitivity of disclosed information. These findings suggest important improvement opportunities for systems design and online communication platforms (Section \ref{implications}), many of which do not take fine-grained interpersonal contexts into account when facilitating cross-cultural communication}\chngnum{14}. 

\MRtext{This paper starts by summarizing related work and then presenting the sociological theories from which we drew to develop each of our specific research questions regarding interpersonal information disclosures (Section~\ref{sec:theory}). We then present our culture-sensitive relationship taxonomy (Section~\ref{sec:data}) which we developed in order to identify relevant tweets. Then we describe our data set which was obtained through large-scale Twitter data collection  (Section~\ref{sec:data_collection}) and \MRtext{dependency-parsing based filtering}. We present our results on culture-specific interpersonal information disclosure norms (Section~\ref{sec:results}) and conclude with implications of this work.}

\section{Related work}\label{sec:rel_work}

\noindent \renewtext{Our work focuses on culture-specific information disclosure online. Information privacy research often focuses on information disclosure as a behavior with key privacy implications. \MRtext{ We build on privacy theories that conceptualize privacy as a negotiation of disclosed information among multiple parties, where information sensitivity is contextually dependent.} Here we give a summary of these theories and provide an overview of literature that \MRtext{investigates interpersonal information disclosure (multi-party privacy)} in the context of social media. }

\textit{Privacy theories.} \newtext{Theories of privacy emphasize the interpersonal and collective nature of privacy management. Altman conceptualizes privacy as a Boundary Regulation where }people balance their privacy needs through a bi-directional interpersonal boundary regulation process across \newtext{different boundaries such as information disclosure, accessibility, territorial, etc. ~\cite{altman_environment_1975} Privacy management is a matter of maintaining the right balance between, e.g., sharing too much and sharing too little, or being overly accessible or too inaccessible.}
Petronio's Communication Privacy Management theory (CPM)~\cite{petronio_boundaries_2002} extends Altman's theory by proposing the interpersonal boundary regulation process as a boundary coordination between the stakeholders of information sharing (i.e., the people who share the information and the people who receive the information). \newtext{In other words, }privacy is an interpersonal process requiring the collective efforts of multiple people \newtext{to ensure that the right balance is maintained for how much information is shared.}
When such coordination fails, boundary turbulence happens, which can raise privacy concerns and drive the stakeholders of the information to resolve the problem and restore the coordination. 
\newtext{We also draw on Nissembaum's Contextual Integrity Framework which focuses on privacy as understanding norms of acceptable transmission principles for information sharing~\cite{nissenbaum-2010-ci-book}. She distinguishes between at least three distinct parties important to establishing the appropriate norm for information sharing: the sender, the recipient, and the subject of the information disclosure. Norms of disclosure for a given set of parties establish what is typical and expected. According to Nissenbaum, a violation of what is expected is when privacy violations occur. In such a way, societies implicitly have an expectation of what is appropriate to share, how, and under what circumstances. However, when considering the context of different cultures such as in different countries, these norms could vary greatly~\cite{solove-2008-privacyBook}.} 

We build on \newtext{these theoretical groundings that define information privacy as a norm-based boundary regulation of information disclosure involving multiple parties. Our study focuses on understanding the culturally-specific norms behind sharing interpersonal information on Twitter. While there is culturally specific research on interpersonal information disclosure observed offline (which we outline in the next section), we are investigating the norms of disclosure for the new information recipient of a public online audience (via Twitter). Understanding the norms of disclosure is a first step towards understanding appropriate disclosures online, which will help us move towards understanding culturally-specific disclosure violations in the future.} 

\newtext{
\textit{Multi-Party Privacy (MPP) in Social Media.}
Some scholars have investigated privacy issues that arise when one discloses information about others on social media. For example, someone can reveal another person's identity, location, or relationship status just by mentioning/tagging them or re-sharing content~\cite{cho_collaborative_2017,cho_networked_2016,hu_detecting_2011,jia_autonomous_2016}. }
A bulk of the MPP literature has focused on group photo sharing. Research has shown that the co-owners involved in group photo sharing often include family, friends, romantic partners, acquaintances, co-workers, and others~\cite{such_photo_2017}. Concerning topics of group photos include people drinking, old photos from the past, children, %
sporting events, concerts, etc.~\cite{such_photo_2017,lampinen_were_2011,ahern_over-exposed_2007}. Sharing photos of others \newtext{raises privacy concerns for those others (co-owners) since} co-owners do not have as much control as the owners over the content~\cite{besmer_moving_2010,fogues_sharing_2017,such_photo_2017,lampinen_were_2011, ahern_over-exposed_2007,pu_valuating_2017}. Co-owners also tend to be concerned about their known social \newtext{circles (e.g., family members, relatives, friends, employers, colleagues)} seeing the photos and misinterpreting them \cite{besmer_moving_2010, lampinen_were_2011}. 

However, little research has \newtext{examined norms of interpersonal information disclosure} in textual information sharing. While textual information sharing does not explicitly reveal the co-owner's image and surroundings, it can reveal other types of important personal information, such as names, activities, and emotions. \newtext{Note that we consider \textit{all} information disclosed about an interpersonal relationship as potentially sensitive since this depends deeply on context and parties involved (e.g., someone may feel that disclosing a step relationship could overshare about their past life events). It is important for us to understand the norms of interpersonal disclosure since privacy violations stem from violating those norms. Thus, by understanding what interpersonal information people are disclosing, how, and with what topics, we are understanding the boundaries of what is acceptable.}

Furthermore, \newtext{few scholars examine MPP in a cross-cultural context,} especially in non-western countries. \newtext{Studies have shown how socio-cultural norms for interaction, self-disclosures, and interpersonal boundaries are different from culture to culture (see Section 3). Thus, we aim to bridge this gap by examining cross-cultural interpersonal information disclosure norms about MPP for text-based content. To do so, we next turn to the extant sociological literature on cultural differences in interpersonal communication which allowed us to focus on facets of information disclosure that have been found to be expressed differently from culture to culture.} %

\section{Cultural Norms of Interpersonal Disclosure}\label{sec:theory}

\renewtext{In this section we draw on the cross-cultural literature that identifies cultural differences in interpersonal communication. 
Culture is a collective concept that describes the "collective programming of the mind that distinguishes the members of one group from others" \cite{hofstede2005cultures}. Researchers have dimensionalized cultural differences into multiple dimensions, such as individualism/collectivism \cite{hofstede2005cultures}, power distance \cite{hofstede2005cultures}, and harmony \cite{schwartz1994beyond}. Communication research has highlighted individualism/collectivism as the most prominent cultural construct to explain the differences in interpersonal communication found between different cultures \cite{cho_qualitative_2013,hsu_cross-cultural_2007,schug_relational_2010,kito_self-disclosure_2005,chen_intercultural_2006,yum_computer-mediated_2005}. This is because individualism/collectivism, compared with other cultural dimensions, is more about relationality \cite{oyserman_rethinking_2002}, which is a central component in interpersonal communication \cite{walther1992interpersonal}. Following this stream of work, we will take a first step towards understanding cultural differences in online communication by focusing on individualism/collectivism and the differences that are expected to result.} 

\renewtext{Reviewing the cross-cultural interpersonal communication literature, we find that the cultural differences in offline interpersonal communication manifest mostly in terms of interpersonal relationships, topic content, and emotional expression. We summarize these three aspects in the following sub-sections:}

\subsection{Interpersonal Relationships Disclosed}
\label{background_rq2}

\renewtext{A key cultural difference found in interpersonal communication is the way people from different cultures talk about their interpersonal relationships. One defining characteristic that drives these differences is the interdependence of relationships. In collectivist cultures, people tend to have high interdependence with their relationships, while in individualistic cultures people have lower interdependence, which influences how they communicate with others.} 

\newtext{More specifically, people have in-group and out-group relationships. People in collectivist cultures form enduring relationships with a few very important in-groups \cite{triandis_individualism_1988}. In collectivist cultures, in-group has been defined as} "family and friends and other people concerned with my welfare" \cite{triandis_analysis_1972}, “parents, friends, neighbors, or coworkers” \cite{hui_measurement_1988}, or “groups with which a real or symbolic blood tie exists (families, tribes, races/ethnicities, religions, nations), peoples as well as civic (neighborhood or community) or other working groups” \cite{reis_collectivism_2009}. \newtext{Collectivist cultures place higher value on their in-group relationships than out-group, thus interact and communicate with in-group relationships more frequently and intimately. Among different in-group relationships, vertical relationships (i.e., parent-child, blood relatives, elders, biological family members) are more important in collectivist} cultures \cite{takahashi_commonalities_2002,realo_hierarchical_1997,shkodriani_individualism_1995}. Interdependence is maximized between parent and child by frequent guidance, consultation, and deep involvement in the child's private life \cite{triandis_individualism_1988}. Extended family members (grandparents, aunts, uncles, and cousins) are also considered to be important in-group relationships in \newtext{collectivist cultures as people are willing to sacrifice personal obligations for those family members }\cite{shkodriani_individualism_1995}. 

\newtext{In contrast, individualistic cultures often perceive their in-group }as "people who are like me in social class, race, beliefs, attitudes, and values."~\cite{triandis_individualism_1988}. Thus, individuals can relate to many in-groups. There is more detachment from in-groups and people have greater skills in entering and leaving new in-groups \cite{triandis_individualism_1988}. Horizontal (i.e., spouse and friend) is the most important in-group relationship in individualistic cultures \cite{takahashi_commonalities_2002}. 

\newtext{Because different relationship types are valued differently in prior cross-cultural studies on interpersonal relationships, it is possible that there is a difference in the frequency of interpersonal information disclosures for tweets from India compared with the U.S. We were thus interested in exploring which relationship types are more commonly disclosed about on Twitter, and whether there is a difference in relative frequency of each disclosed relationship type with respect to the others, both across cultures and within cultures:}

\newtext{\textit{RQ1: How does the frequency of disclosure about different interpersonal relationships (family, friends, co-workers, etc.) on Twitter differ between India and the U.S?}}

\subsection{Topics Disclosed}\label{background_rq1}
\noindent \newtext{We also reviewed the content and topics of interpersonal disclosure identified in past studies.
Prior work has shown that people in individualistic cultures tend to \textit{self-disclose} more than in collectivist cultures \cite{cho_qualitative_2013,hsu_cross-cultural_2007,schug_relational_2010}. This observation holds across different types of interpersonal relationships. Several studies of collectivist cultures even report lower levels of self-disclosure }with in-group relationships, such as romantic relationships and friends \cite{kito_self-disclosure_2005,chen_intercultural_2006}. \newtext{In fact, in East Asian cultures,} if one person reveals too much about himself or herself, the other may take it as inappropriate or as an indicator of incompetence \cite{yum_computer-mediated_2005}. \newtext{On the other hand, a positive association} of self-disclosure and intimacy/trust has emerged when studying communication in individualistic cultures \cite{yum_computer-mediated_2005}.

Gudykunst and Nishida \cite{cudykunst_social_1983} further find that people in individualistic countries tend to talk more about their marriage, family, relationship with others, love/dating and sex, emotions and feelings, interests/hobbies, and their attitudes/values. Those in collectivist countries, instead, talk more about physical condition, school/work, biographical information, religion, and money/property. \newtext{Similarly, Cahn \cite{cushman_communication_1985} found that people in individualistic countries discussed more intimate topics while people in collectivist countries discussed more superficial topics. Chen \cite{chen_differences_1995} found that individualistic cultures disclose more about topics of opinions, interests, work, financial issues, personality, and body than collectivist cultures.}

\newtext{Motivated by these findings about topical differences in offline interpersonal disclosures, we will explore whether such cross-cultural differences also exist in the topics disclosed online. Therefore, we are interested in answering following sub-question:}

\newtext{
\textit{RQ2: How do the topics of interpersonal information disclosures on Twitter differ between India and U.S.?}}

\newtext{Moreover, prior research has shown that people in individualistic countries are more concerned with and are less likely to disclose sensitive personal information on social media than those in collectivist countries \cite{chen_differences_1995,cho_qualitative_2013,posey_proposing_2010,tsoi_privacy_2011}.}
\newtext{One reason is that users in collectivist countries primarily use social media to maintain their current relationships, especially close ties and offline connections belonging to the same social groups \cite{kim2011cultural, tsoi_privacy_2011}. Thus, they tend to reveal more personal details in online disclosure \cite{peters2015cultural}. On the other hand, users in individualistic countries usually have a wider variety of online social networks on social media  \cite{kim2011cultural}, thus adopt a more protective means of self-disclosure \cite{rui2013strategic}. In light of this, we focus on two types of information that have been shown to be sensitive in many studies - specifically, location and financial information \cite{tsai2010location, wsjlocation}.}

\newtext{In following with prior research, we expect that sensitive personal information is differently disclosed by those from an individualist culture than by those from a collectivist one. We thus will explore the following sub-question regarding how interpersonal disclosure norms for information that is \MRtext{potentially} sensitive will differ between India and U.S. tweets (Results presented in~\ref{findings_rq4}):}

\newtext{\textit{RQ3: How does \MRtext{potentially} sensitive information, such as location and financial information, in interpersonal information disclosure on Twitter differ between India and U.S.?}}  

\subsection{Emotional Disclosure Norms}
\label{background_rq3}
People in individualistic cultures tend to express their emotions more \newtext{ directly}, compared to those in collectivist cultures \cite{takahashi_commonalities_2002,branz_sentiment_2018}. Overall, they tend to elaborate, highlight, or emphasize positive feelings much more than negative feelings \cite{kitayama_culture_2000,branz_sentiment_2018}. However, such tendency \newtext{can be moderated by the specific interpersonal relationships with which the person is interacting}. Many cross-cultural studies have shown that people in collectivist cultures tend to \newtext{express more positive emotions and less negative emotion towards their \textit{in-group} relationships (e.g., family, friends, other people concerned with their welfare), }whereas people in individualistic cultures do not show different tendencies of positive/negative emotion \newtext{ towards in-group versus out-groups \cite{kitayama_cultural_2006,matsumoto_mapping_2008}. They still clearly communicate negative emotions towards in-group relations} when needed \cite{ho_component_1994,markus_culture_nodate,triandis_individualism_1995}. 
\newtext{This traces back to the value placed on in-group harmony by those in collectivist cultures \cite{triandis_individualism_1988,takahashi_commonalities_2002}. To maintain harmony, communicators are sensitive to conveying and receiving contextual cues including indications of }belonging, dependency, empathy, norms for reciprocity, and occupying an appropriate place within the relationship hierarchy \cite{reis_collectivism_2009}. \newtext{This makes it important to effectively regulate one's emotional displays so as not to threaten in-group harmony and cohesion. For example,} positive emotion brings people together whereas negative emotion drives people apart \cite{matsumoto_mapping_2008,triandis_individualism_1988}. Thus, \newtext{speakers from collectivist cultures tend to use positive expression towards in-group relationships. However, they create greater distance between in-groups and out-groups by using more negative emotional expression toward out-group members} \cite{matsumoto_mapping_2008}. 

On the \newtext{other end of the spectrum, speakers from individualistic cultures value independence and individual goals. This is reflected in their patterns of communication where free expression of emotions reaffirms the independence of the individual, leading to an overall greater amount of emotional expression \cite{matsumoto_mapping_2008}. The variety of emotional expression may also be greater in individualistic cultures, which also reflects the value of individualistic expression \cite{matsumoto_mapping_2008}. The emotive characteristics of communications about in-group and out-group relationships do not differ significantly and are more direct, which facilitates smooth interaction with strangers \cite{oyserman_rethinking_2002}. Indeed, little difference has been observed in emotional expression towards }in-group and out-groups \cite{matsumoto_mapping_2008}.

\newtext{Drawing from this earlier cross-cultural communications research, we argue that the frequency and the polarity of emotional expression in interpersonal information disclosures are likely to differ between individualistic and collectivist cultures. We will thus explore the following sub-questions regarding the emotion frequency and polarity in interpersonal information disclosures online:} %

\textit{RQ4: In the context of interpersonal disclosure, how does the frequency of emotions disclosed in tweets differ between India and the U.S.?} 

\textit{RQ5: In the context of interpersonal disclosure, how does the frequency of positive/negative emotions disclosed in tweets differ between India and the U.S.?}

\newtext{To answer these research questions, we first had to identify the tweets which disclose information about one's interpersonal relationships. The first step in doing so was to create a novel culturally-sensitive \textit{saturated} taxonomy of keywords that represent interpersonal relationships. Leveraging this taxonomy allowed us to extract tweets potentially containing interpersonal disclosures }from large-scale Twitter data for further exploration. We next present our approach to developing the taxonomy.

\section{Developing a Culturally-sensitive Taxonomy of Interpersonal Relationships}\label{sec:data}

\noindent \newtext{The goal of this study is to understand and compare India and U.S. disclosure norms when people talk about their interpersonal relationships. In order to compare disclosures about interpersonal relationships, we needed to first understand which relationship keywords are perceived as the same type of relationship within a given culture. }

Prior studies have classified relationships based on Fiske’s Relational Model~\cite{fiske_structures_1991} that uses length of interaction, duration of interaction, and intimacy/comfort to differentiate relationships into kinship versus non-kinship relationships~\cite{ho-1998-kinship}. Other research has used participant labeling to categorize relationships into various groups (family, friend, acquaintance, etc.)~\cite{duck-1991-relationshipcat, avrahami_responsiveness_2006}. 
However, such studies do not address how cultural differences might influence which relationships are grouped together. For instance, a member from an Eastern collectivist culture (India) may have a broader, more encompassing perception of family, i.e., they may include aunts, uncles, cousins, etc. in their definition of family. On the other hand, a person from a western individualistic society (U.S.) may have a much narrower definition of family comprising only of parents and their children. In order to gain better insight into the cultural contexts through which relationships are clustered, and thus enable comparison across cultures, we designed a card-sort study to probe on how individuals in different cultures cluster their interpersonal relationships.  

\subsection{Card-sort Study Design and Participants}\label{sec:card-sort-design}

\begin{figure}[!tbh]
\centering
\includegraphics[width=1\linewidth]{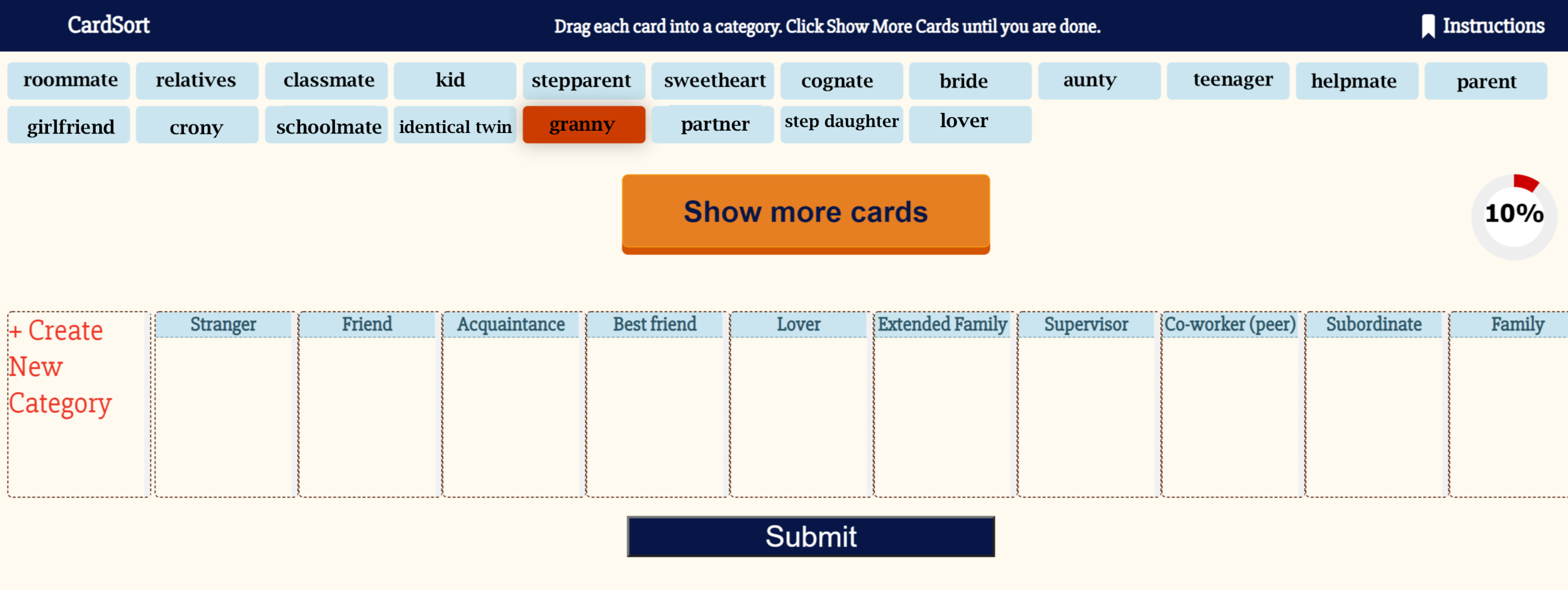}
\caption{A screenshot of the card-sort interface we developed to create saturated relationship taxonomy.}
\label{fig:picture1}
\end{figure}

\noindent The card sort study was designed to be completed in a fully online setting so that 1) We could get a diverse set of participants from geographically dispersed locations, and 2) We could comply with the safety guidelines necessary in the current pandemic. The card-sort task had two main goals: 1) To create a saturated list of keywords that represent interpersonal relationships, and 2) To build a culturally-sensitive taxonomy of relationship types. A card-sort task is a method that allows participants to sort words into groups based on affinity. We used a hybrid card sort. Drawing from a closed sort approach, users were given a predefined set of relationship categories derived from the literature (e.g., family, extended family, superior) as well as a set of relationship words to categorize. We supplemented this with an open sort approach of letting users also create their own relationship categories or words they felt were missing. This hybrid method allowed us to test the exhaustiveness of our initial set of relationship words and relationship groups and enable participants to augment these.

The initial list of words was created by having authors that are natives of the U.S. (N=3) and of India (N=2) generate lists of words used to represent interpersonal relationships. We then performed a dictionary and thesaurus search on those words, repeating the process on new words found, and iterating until no more new words could be found. This resulted in a list of 177 relationship words.

An evaluation of commercially available card-sort tools revealed that they were mostly designed for open or closed card sort studies, but not hybrid. We therefore designed and built an interface that would allow us to conduct a hybrid card sort study. A screenshot of the interface is shown in Fig. \ref{fig:picture1}. Participants were shown instructions to drag and drop each relationship word into any one of the 10 predefined relationship categories (stranger, friend, acquaintance, best friend, lover, extended family, supervisor, co-worker, subordinate, and family) as well as how to create new categories. Each word could only be placed in a single category. Once participants had finished sorting the first set of words, they clicked the ``Show more cards'' button to see the next set of words. Once participants had finished sorting all of the words, they were prompted to create any additional words. This web-based card sort program was piloted for usability before deploying to participants. Our study protocol was approved (before deployment) by the IRB board of the last author's institution. 

\textbf{Participants.} We leveraged the crowd-sourcing platform Amazon Mechanical Turk (AMT) to enroll a total of 121 participants (63 U.S., 58 India). All of our participants had $>$95\% approval rating on AMT, were aged 18 years and older, and spoke English. They completed the card sorting task in 4.5 minutes on average and were compensated \$4.00 for their task. 52.4\% of the U.S. participants were female, whereas 67.3\% of the Indian participants were female. For both cultures a majority (93\% for India and 63.5\% for the U.S.) were between 25 to 44 years old. Fig.~\ref{fig:demogcardSort} in Appendix~\ref{sec:demog} contains a breakdown of our participant demographics.

\subsection{Developing A Saturated Taxonomy of Online Interpersonal Relationships}
\label{sec:taxonomy}
\label{sec:lexicon}

\newtext{Upon carrying out the card sort study, we analyzed the data to create culture-specific lists }of relationship words. We first identified any new words that had been added by participants. Any words that were repetitions, typos (e.g., wefe), or not associated with specific relationships (e.g., their, each other) were excluded from the list of words. For the remaining words we included ones that at least two people had added. The final list consisted of 179 Indian words and 178 U.S. words that participants identified as being relationship words.

\newtext{Next we identified the relationship categories that should be used to group these words, which are necessary to be able to compare types of relationships across cultures. The card sort task was seeded with ten initial relationship categories where participants could group }similar relationship words. However, participants could also create their own categories as needed to classify words. We performed a similarity analysis to identify \newtext{which relationship words should been placed in each category. Namely, if at least} 80\% of participants placed a particular word in a category (e.g., `Mom' placed in `Family') that word was considered associated with that relationship category. Out of the total 178 relationship words there were 81 words in U.S. and 112 words in India that had a clear majority grouping. \newtext{These relationship categories are the ones we used in the final taxonomy.}

\begin{table}
    \centering
    \small
    \begin{tabular}{p{5em} p{3em} p{32 em}}
    Category & Country & Relationship words \\
    \toprule
    Acquaintance & India & crony, folk, folks, progenitor, progeniture, neighbour \\
    & U.S. & clan, clansperson, classmate, schoolmate, neighbour\\
    Supervisor & India & advisor, boss, guru, manager, master, mentor, senior, supervisor, teacher\\
    & U.S. & advisor, boss, guru, manager, master, mentor, senior, supervisor, teacher\\
    Co-worker & India & cohort, colleague, coworker, co-worker, helpmate, partner, teammate\\
    & U.S. & associate, colleague, coworker, co-worker, teammate \\
    Subordinate & India & associate, junior, mentee, subordinate\\
    & U.S. & mentee, subordinate \\
    Friend & India & bosom buddy, bro, buddy, chum, classmate, friend, friends, pal, playmate, roomie, roommate, schoolmate\\
    & U.S. & buddy, chum, crony, friend, friends, mate, pal, playmate, roomie, roommate\\
    Best friend & India & dearest, mate, sidekick, bestie \\
    & U.S. & bosom buddy, sidekick, bestie\\
    Lover & India & babe, boyfriend, bride, bridegroom, darling, fiancee, girlfriend, honey, hun, love, love of my life, lover, soulmate, sweetheart, sweety, true love\\
    & U.S. & babe, better half, boyfriend, bride, bridegroom, consort, darling, dearest, fiancee, girlfriend, honey, hubby, hun, husband, love, love of my life, lover, menage, partner, soul mate, spouse, sweetheart, sweety, true love, wife\\
    Family & India & aunt, auntie, aunty, baby, better half, blood relatives, brother, brotherly, child, children, close-knit, consort, cousin, dad, daddy, daughter, daughter-in-law, elder brother, elder sister, fam, family, fraternal, father, father-in-law, grampa, grandchild, grandchildren, granddaughter, grandfather, grandma, grandmother, grandpa, grandparent, grandparents, grandson, granny, great aunt, great granddaughter, great grandfather, great grandmother, great grandparents, great grandson, great uncle, guardian, hubby, husband, in-law, in-laws, kid, kids, ma, mama, maternal, mom, mommy, mother, mother-in-law, mum, mummy, nephew, niece, offspring, pa, papa, parent, paternal, sib, sibling, siblings, sis, sister, sister-in-law, sisterhood, sisterly, son, son-in-law, spouse, toddler, twin brother, twin sister, uncle, wife, younger brother, younger sister\\
    & U.S. & adoptive father, adoptive mother, baby, bro, brother, brotherly, child, children, close-knit, dad, daddy, daughter, elder brother, elder sister, fam, family, fraternal, father, grampa, grandchild, grandchildren, granddaughter, grandfather, grandma, grandmother, grandpa, grandparent, grandparents, grandson, granny, guardian, half brother, half sister, infant, family, kid, kids, ma, mama, maternal, mom, mommy, mother, mum, mummy, offspring, pa, papa, parent, paternal, sib, sibling, siblings, sis, sister, sisterhood, sisterly, son, step brother, step father, step mother, step sis, step sister, stepbro, stepchild, stepchildren, stepdad, stepmom, stepparent, stepson, toddler, twin brother, twin sister, younger brother, younger sister
\\
    Extended family & India & adoptive father, adoptive mother, ancestor, clan, clansperson, descendant, distant relatives, extended family, forebears, half brother, half sister, infant, kin, kindred, kinfolk, kinship, kinsperson, kith, progeny, relative, relatives, step brother, step daughter, step father, step mother, step sis, step sister, stepbro, stepchild, stepchildren, stepdad, stepmom, stepparent, stepson\\
     & U.S. & ancestor, aunt, auntie, aunty, blood relatives, brother-in-law, cousin, daughter-in-law, descendant, distant relatives, extended family, father-in-law, folk, folks, forebears, great aunt, great granddaughter, great grandfather, great grandmother, great grandparents, great grandson, great uncle, in-law, in-laws, kindred, kinfolk, kinship, kinsperson, mother-in-law, nephew, niece, relative, relatives, sister-in-law, son-in-law, step daughter, uncle, progenitor, progeniture, progeny, kin
\\
    \bottomrule
    \end{tabular}
   \caption{\newtext{Final relationship groupings by country based on participant responses.}} \label{tab:final_relationship_groupings}
\end{table}

Final groupings are shown in Table~\ref{tab:final_relationship_groupings}.
\newtext{The word lists and relationship categories identified in this card-sort study are both \textbf{saturated}, meaning participants were able to add words and relationship categories until they felt these lists were complete, and \textbf{culture-specific}, meaning that they represent word groupings with broad agreement of at least 80\% of the members of the given culture.} For example, ``uncle'' belongs to the \textit{Family} category in India, but to the \textit{Extended Family} category in the U.S., while ``wife'' and `husband'' belong to the \textit{Lover} category in the U.S. and to the \textit{Family} category in India. \newtext{This allows us the ability to} analyze disclosure patterns according to the roles individuals are assigned in a given culture.

\section{Developing a Culture-sensitive Interpersonal Disclosure Tweet Corpus}
\label{sec:data_collection}

\renewtext{In accordance with existing literature on multi-party privacy, we seek to identify tweets that disclose information about a third party with whom the tweeter has an existing interpersonal relationship. Such disclosed information may potentially constitute a privacy violation in one culture even if the disclosed information appears innocuous in another culture. To study this phenomenon, we
leverage the taxonomy developed in Section \ref{sec:data} to collect a dataset of 417,953 U.S. and 33,591
India tweets containing interpersonal relationship disclosures.}

\subsection{Collecting Large-scale Twitter Data Using Our Taxonomy} 

\newtext{To collect the tweets, we used a Twitter data collection tool called Twint~\cite{twint}, which allows the extraction of past tweets based on keywords. We set geographic perimeters to specifically collect tweets from within India and the U.S. The implicit assumption in this choice is that tweets originating within India will be generally aligned with Indian cultural norms while tweets originating within the U.S. will be more aligned with continental U.S. norms. Naturally, it is possible that expatriate visitors to each country may live in one region while adhering to the social norms of another, however, we rely on the assumption that an expatriate living in foreign country likely exhibits behaviors that are shifted toward the prevailing norms of the host country, even if they do not fully conform to those norms. Thus, expatriates might dilute our results somewhat, but do not invalidate them. }\chngnum{15}
A very recent work also explored potential biased in Twint collected data (as compared to Twitter's API provided random 1\% sample data)~\cite{poddar22}. This work demonstrated that for a set of keywords and a specified time window, Twint provides a random sample around 15\% of all tweets containing those keywords and posted in the time window. Furthermore, Twint provided data is representative both with respect to time and tweet popularity, \newtext{giving us assurance that our collected data is unlikely to have systematic bias in this regards.}

\newtext{We used Twint to collect tweets from three distinct time periods. We chose the time periods by considering the following criteria: (a) We wished to avoid any data produced during the global coronavirus pandemic, (b) We desired the tweets to be as recent as possible, (c) We wanted a somewhat representative spread of data (i.e., not all from the same holiday season). Based on these criteria, we selected August 2019, December 2019, and January 2020 for data collection. December 2019 and January 2020, the final two months prior to the global coronavirus outbreak, represent the most recent available tweet data that was not influenced by the widespread emotional impact of the virus. To balance the data set and prevent undue influence from end-of-year holiday seasons, we also collected tweets from Summer 2019. These tweets were produced during a time with different weather patterns and was the most recent month without major holidays. The overall intent was to approximate a generic time frame.}\chngnum{2}

\newtext{We collected tweets from the aforementioned time periods that contained the relationship words identified}\chngnum{2} in the lexicon from Section~\ref{sec:lexicon}, resulting in about 4.5 million tweets. The Twint extraction process creates the possibility of extracting the \newtext{exact same tweet with the same unique tweet id}\chngnum{3} %
\newtext{multiple times (for example, due to existence of multiple keywords in a tweet),}\chngnum{3} 
so we undertook a data cleaning process to excluded such duplicate tweets. We also removed urls, user mentions (e.g., @cscw), and punctuation, and then converted each tweet to lower case. Finally, we kept only English words in each tweet which allowed us to make direct comparisons in our analysis.
After the data cleaning process, we had a total of 2,095,792 U.S. and 272,755 Indian tweets originating from 604,895 unique users.

\begin{center}
\begin{table}[h!]
\small
\begin{tabular}{|m{9cm}|m{2cm}|m{1.7cm}|}
 \hline
 \textbf{Tweet} & \textbf{Relationship Word} & \textbf{Interpersonal Disclosure} \\
 \hline
 My boss asking me to multitask department, because we short of people \& very stingy to hire another & boss & YES \\
 \hline
Been in mourning these days through all the happenings in the world... Especially the lost of a real one for the culture... Kobe and his daughter. May they rest in peace forever! My next jump shot is for… & daughter & NO \\
 \hline
 My daughter’s health has been one of my biggest worries. She was a premature baby and I have always been concerned about her nutritional needs. Recently, I found out that milk adulteration is BIG problem in India.… & daughter & YES \\
 \hline 
 ``One and only 4 Dhanush Fan'' Say no to others Hero..Confident boss confident boss..Our Thalaivar $@$[user name] history's say everything & boss & NO \\ 
\hline
\end{tabular}
\caption{Examples of tweets with relationship words and and whether they refer to an interpersonal disclosure}
\label{table:valid_invalid}
\end{table}
\end{center}

Manual inspection of the resulting dataset revealed that while all of the web-scraped tweets contained at least one word from our relationship lexicon (Section~\ref{sec:lexicon}), 
these words were not always used to indicate a direct interpersonal relationship between the speaker and the referenced entity. Common sources of confusion were references to public figures or prominent news stories, and using words as homonyms in a non-relationship context \newtext{(see examples in }Table~\ref{table:valid_invalid}). To filter out these invalid tweets, we implemented a dependency-parsing based classification algorithm \newtext{which we describe next.} 

\subsection{Identifying Relevant Tweets}
\label{sec:identifying_relevant_tweets}

In order to identify tweets that refer to a valid interpersonal relationship (as opposed to merely containing a relationship keyword), we designed a classifier using the dependency parsing tools in the spacy python package \cite{spacy}. The classifier was premised on the observation that most valid relationship tweets included a syntactic connection between the referenced relationship and the tweeter, e.g. "\textit{my} boss", "\textit{our} long-lost aunt". Accordingly, we designed the classifier to accept only tweets that contained a linguistic dependency between the relationship word and any one of {`me', `my', `our'}. This method might miss some valid tweets, but our study goal prioritizes precision over recall (i.e., to ensure that we conservatively include only valid tweets).

To confirm that the method was working correctly, \newtext{two members of our research team manually (and independently) labeled a random sample of} 1,240 tweets that had been labeled by the classifier. Using these manual labels as ground truth, the precision for valid tweets was 0.92 and recall was 0.81, with a corresponding F1 score of 0.86. These scores indicated that the classifier was performing well and could be used on the full dataset. We further constrained each tweet to be classified as belonging to one and only one relationship category. After this more stringent dependency-based classifier had been applied, we obtained
417,953 U.S. and 33,591
India tweets that had been classified as containing valid interpersonal relationship disclosures. 

\subsection{Limitations \MRtext{and Ethical Considerations}}\label{sec:limitations}

\noindent Like any empirical study, our data-driven analysis of cultural norms for interpersonal disclosure also is limited by our dataset. \MRtext{Here we identify several potential limitations of our approach as well as discuss ethical considerations in choosing this dataset. }

We primarily considered only English relationship keywords from our taxonomy to find interpersonal relationships from Twitter in our dataset---this is quite appropriate for users from the U.S. but might have resulted in an incomplete set of norms for Indian users (since Indian users can use many non-English languages). We feel our data set is still very appropriate for our study---(1) Using only English keywords allowed us to more directly compare relationship-related disclosures between India and the U.S. (future work should investigate translating from other languages). (2) We note that the English-speaking Twitter users from India are potentially already familiar with at least some norms of individualistic societies like the U.S. Thus, the cultural norm comparisons reported in the paper may serve as a conservative estimate of actual norm differences between the U.S. and India, underscoring the external validity of our study. 

We note that word ordering in Indian tweets may be different from U.S. tweets. Consequently, given our machine learning and deep learning algorithms are often trained on U.S. English data (e.g., for dependency parsing and emotion detection), their results might be invalid for Indian tweets (although written in English). We addressed this limitation by randomly sampling a set of U.S. and Indian tweets after each application of such machine learning techniques and manually checking their labels. Unless otherwise stated, all of our learning models yield valid results (as manually confirmed on our random samples) on both tweets from India as well as the U.S. 

We focused on analyzing an initial set of emotions and \MRtext{data} (location and financial). Our general qualitative analysis of the content also allowed us to capture a wider array of characteristics and information types. However, future work could focus on analysis of specific additional types of emotions and data.

By analyzing a specific online social network, Twitter, it may provide a partial picture of culture-specific norms of online disclosure. However, the international popularity of the platform allowed us to collect data from the two cultures which already gives some initial insight into interpersonal disclosure norms. \newtext{We also were able to ground our findings in a real-world context and uncover the norms for a given platform.} Future work should expand to additional data sources. 

We note that we collected tweets after-the-fact and it's possible that there were tweets deleted between time-to-post and time-to-collect~\cite{mondal16}. However, \MRtext{by taking this approach we respect the implicit privacy wishes expressed by the posters in deleting the post. Furthermore, this does not present a major issue to our study for being able to identify cultural norms}. The data set of tweets left are the ones users are comfortable leaving on Twitter, which speaks to the norms of what is an acceptable disclosure. 

\MRtext{We focus on location and financial information where certain types of this information has been found to be sensitive in prior research. For example, precise detailed GPS coordinates is sensitive when shared to certain audiences in certain contexts~\cite{karen-2006-privlocation-data}. However, it is not necessarily sensitive when shared in other contexts with other audiences~\cite{Page_Kobsa_Knijnenburg_2021}, or at different levels of granularity or with different types of data representation such as being at ``home'' versus ``work''~\cite{whereabouts-brown-2007-ubicomp}. Thus, while we do identify location and financial data in our dataset, we are focusing on the types of financial and location data that are appropriate to share publicly rather than looking at more sensitive contexts such as private tweets. As noted previously, any tweets that the user decided were not appropriate could be deleted and would not be included in this data set. Nonetheless, we add an additional layer of deanonymization by removing the usernames from the tweets. Furthermore, the tweets are aggregated for our analysis and reporting of results. Additionally, all of our protocols underwent IRB review and were approved, with the Twitter data collection and analysis deemed exempt and considered minimal risk. To further examine the ethical aspects of our data collection from Twitter, we referred to the work of Eysenbach and Till~\cite{eysenbach-2001-ethics} who suggested three properties to decide if a forum is public---(i) Is registration necessary to view a post (the answer for Twitter is no). (ii) size of the forum which for Twitter is in hundreds of millions (iii) how members perceived this forum. While it's difficult to tell how our specific posters felt, Twitter is treated as a public forum widely cited by politicians, celebrities, and the media (e.g., during the Indian General election, during COVID) ~\cite{IJoC6705, PAL201797, venigalla-2020-covid-twitter-india}. This reinforces the public nature of the forum. }

\MRtext{Overall, we believe we have taken measures to minimize any risks associated with doing such research}. We now turn to the results of our study.

\section{Results}\label{sec:results}

\subsection{Personal relationships within and across cultures (RQ1)}
\label{findings_rq2}
\begin{figure}
\centering
\includegraphics[width=0.6\linewidth]{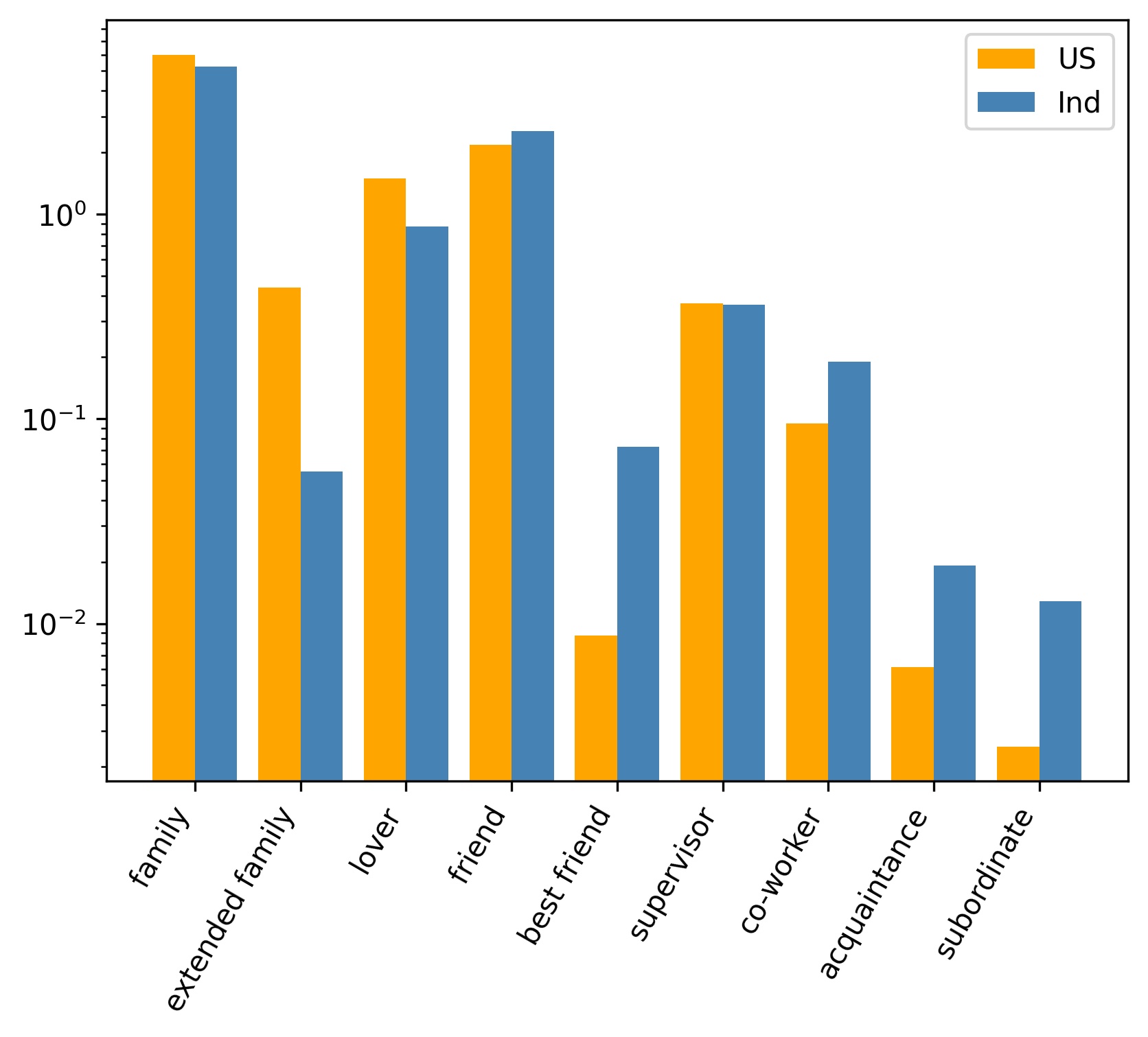}
\caption{\newtext{Percentage-wise breakdown of 417,953 U.S. tweets and 33,591 India tweets referencing valid interpersonal relationships. US tweets are shown in \MRtext{orange}, India tweets in \MRtext{blue}. \MRtext{The scale is logarithmic.}}
}
\label{fig:dataset_all}
\end{figure}

\newtext{We begin by examining the frequency with which members of each culture invoke direct interpersonal relationships (e.g., `\textit{my} co-worker', `\textit{our} daughter') as opposed to referencing relationships in general (e.g. `a good co-worker is always on time', or `daughters are a delight'). We were able to do this using the tweet data set described in Section \ref{sec:data_collection} and the dependency-parsing filter described in Section \ref{sec:identifying_relevant_tweets}, as shown in Figure \ref{fig:dataset_all}. We find that U.S. and Indian disclosure patterns differ across all relationship categories, with the most dramatic differences occurring in the Extended Family and Lover categories. U.S. tweeters are 7.3 times and 1.7 times more likely, respectively, than Indian tweeters to reference these relationships as compared to other types of interpersonal disclosures. Indian tweeters, in contrast, are 7.8 times more likely to disclose information about best friends and 1.9 times more likely to mention co-workers}

\newtext{It is important to note that this result illustrates the \textit{relative} importance (as indicated by frequency) of a given relationship in comparison with other relationship types, but not the overall frequency of interpersonal disclosures generally. This work examined the (relatively small) subset of all tweets that included interpersonal references. We are therefore unable to discuss how frequently Indian vs U.S. tweeters refer to interpersonal relationships overall. Instead, we can observe that Twitter users in both cultures prefer to tweet about Family and Friend relationships over more distant relationships such as Acquaintance, Co-worker, Extended Family, and Supervisor. We are also able to observe that when considering frequencies \textit{across} cultures, U.S. tweeters place greater relative emphasis on the Lover, Family and Extended Family categories, while Indian tweeters place greater relative emphasis on Friend, Best Friend, and Co-worker relationships.}

\newtext{These findings run counter to many assumptions about collectivist vs. individualist cultures. For example, both cultures show strong tendencies to refer to strong in-group relations (Family, Extended Family, Friend/Best Friend, Lover) and lower tendencies to refer to less integral relationships (Supervisor, Co-worker, Acquaintance, Subordinate). While we would expect a collectivist culture (India) to place higher relative importance on Family and Extended Family relationships, we find instead that U.S. tweeters are more likely to reference those groups, while Indian tweeters outpace their U.S. counterparts in disclosing information about co-worker and best friend relationships. This can be explained in part by the culture-specific nature of our card-sort taxonomy. The Lover category, for U.S. tweeters, includes not only non-institutionalized romantic relationships but also spouse relationships such as husband and wife, which may account for its greater representation within U.S. tweets. On the other hand, the Family category for U.S. tweeters contains only 75 specific relationship words as compared to India's 84, and yet U.S. tweeters refer to this category with higher frequency than do tweeters based in India.}

\newtext{Overall, we find that while large-scale trends in relationship frequency follow similar patterns across both countries, there are important and intriguing differences in many relationship categories.}

\subsection{Topical Norms of Information Disclosure Across Cultures (RQ2)}
\label{findings_rq1}

\newtext{In this section, we seek to explore the broad topics of discussion introduced by U.S. vs. Indian tweeters when disclosing information about third parties with whom they have a direct relationship.}

\subsubsection{Discovering topics using Latent Dirichlet Allocation (LDA) and manual coding} Since we wanted to understand the broad and general topics disclosed by users about their interpersonal relationships, we leveraged an unsupervised method. Specifically, we used LDA \cite{blei-2003-lda}, a widely-used method to find major themes of conversation from a collection of tweets. LDA takes a set of texts and the number of topics as input and outputs clusters of words. Each cluster identifies a topic along with the probability of each tweet belonging to that topic. 

\MRtext{This work used the gensim LDA package \cite{gensim2011} with parameters and setup similar to those used by ~\citet{budak_what_2019}. We used a learning decay rate of 0.5 across 20 passes with 400 iterations in each pass, using alpha and eta values as automatically determined by the gensim package. Chunk size was set to 30,000.} We pre-processed each U.S. tweet by first removing the relationship keywords as well as stopwords to facilitate extraction of the meaningful phrases containing disclosed information about specific relationships. We then computed the topic coherence scores of our LDA model while varying the number of topics from two to forty; We observed that the coherence scores were saturated for LDA models within 40 topics~\cite{budak_what_2019}. Finally, we chose the least number of topics which resulted in the highest coherence score for our U.S. specific LDA model. We repeated the same steps for the corpus of Indian tweets. In total we auto-identified 18 topics (word clusters) for India and 26 topics for the U.S. Next, two researchers (one from India and another from the U.S.) collaboratively went over each of the topics (clusters of words) as well as 10 tweets that are most likely to be part of that topic. They assigned a human-readable theme to each topic of each country (based on the word clusters as well as tweets). Furthermore, they collaboratively collapsed the topics which shared a similar broader theme~\cite{braun_thematic_2019}. At the end of this thematic analysis we had a total of 21 topics (12 for U.S. and 9 for India) which characterizes the interpersonal topical information disclosure. We present the final topics in Table~\ref{table:full_topic_disclosure_comparison}. Additionally, the details of the word clusters for all the topics are in Appendix~\ref{sec:extra_topical_disclosure}. 

\begin{table}
	\small
    \begin{tabular}{p{10em}|p{6em}||p{10em}|p{6em}}
    	\multicolumn{2}{c}{U.S.} & \multicolumn{2}{c}{India}\\
    	\hline
        Topics Disclosed  & \% Tweets  & Topics Disclosed  & \% Tweets\\
        \toprule
        \cellcolor[gray]{0.8}Stories about Family & \cellcolor[gray]{0.8}35.5\%  & \cellcolor[gray]{0.8}Stories about Family & \cellcolor[gray]{0.8}26.1\%  \\ 
        \cellcolor[gray]{0.8}Complaining & \cellcolor[gray]{0.8}10.6\% &  \cellcolor[gray]{0.8}Celebrations & \cellcolor[gray]{0.8}25.9\% \\ 
        Gratitude & 10.4\%  & Expressing Love  & 19.5\% \\ 
        \cellcolor[gray]{0.8}Celebrations & \cellcolor[gray]{0.8}9.8\%  & Patriotism  & 9.7\%  \\ 
        Christianity & 8.9\%   & \cellcolor[gray]{0.8}Work & \cellcolor[gray]{0.8}5.7\%  \\ 
        Politics & 6.8\% &  \cellcolor[gray]{0.8}Schooling  & \cellcolor[gray]{0.8}4.7\% \\ 
        Profane Narrative & 5.9\%  & \cellcolor[gray]{0.8}Reminiscing  & \cellcolor[gray]{0.8}3.4\% \\ 
        Female Romantic partner & 5.0\%  & \cellcolor[gray]{0.8}Complaining  & \cellcolor[gray]{0.8}2.9\% \\ 
        \cellcolor[gray]{0.8}Schooling & \cellcolor[gray]{0.8}3.8\% & Other & 2.0\% \\
        Social Media Activities & 1.6\% & & \\ 
        \cellcolor[gray]{0.8}Work & \cellcolor[gray]{0.8}0.9\% & & \\
        \cellcolor[gray]{0.8}Reminiscing  & \cellcolor[gray]{0.8}0.8\% & & \\ 
        \bottomrule
    \end{tabular}
    \caption{Topics revealed by users in their information disclosures about their interpersonal relationships. Six topics are common across cultures: 'Celebrations', 'Work', 'Complaining', 'Reminiscing', 'Schooling', 'Stories about Family'. However, some topics are specific to different cultures, e.g., 'Patriotism' and 
    Expressing love' seem to be particular to the disclosure norms of India, while U.S. users disclosed six extra topics (not present in India): 'Gratitude', 'Christianity', 'Female Romantic partner', 'Politics', 'Profane Narrative', and 'Social Media Activities'. Upon further investigation, we uncovered that the `Expressive love' topic for Indian users often corresponds to love expressed towards kids, best friends, public figures/idols, and almost never about their partner. \MRtext{We marked the \colorbox[gray]{0.8}{six common topics in gray} for ease of reading.}}
    \label{table:full_topic_disclosure_comparison}
\end{table}

\subsubsection{Identifying themes of topical disclosure}
For each of the disclosed topics, we sampled 40 tweets (20 from each culture) \MRtext{that had passed the relevance tests described in \ref{sec:identifying_relevant_tweets}} for further analysis. Two researchers went over both sets of 20 tweets from each topic and performed thematic analysis to uncover \textit{why} and \textit{about whom} the users are largely using a topic~\cite{braun_thematic_2019}. This analysis identified how U.S. and Indian users behave similarly while disclosing about a few topics, but very differently when talking about the rest. 

\subsubsection{Thematic similarities of topical disclosure norms between India and the U.S.}
Twitter users from both countries disclose information on five similar topics: \textit{Schooling, Stories about their families, Holidays/celebrations, Reminiscing, and Their daily work life}. The schooling category is occupied primarily by tweets about primary and secondary education, with some comments speaking about college life. In both countries, they tweet about parents, children, and siblings, and their participation in the educational system and process. They also keep a narrative and informal tone when telling stories about family, focusing on interesting details, entertainment, events attended, or food. While holidays are different, the ways that Twitter users from both countries communicate about them are similar, often focusing on statements of gratitude for their families. When it comes to celebrations, the U.S. tweets are primarily about birthdays and Christmas, whereas (perhaps naturally) the India tweets reference a wider variety of holidays. 

\subsubsection{Thematic differences of topical disclosure norms between India and the U.S}
We observed that although complaining is a common topic across both cultures, Indians complained about the services of organizations. In contrast, U.S. users tend to complain more about their family members. However, U.S. tweets often expressed appreciation for spouses (which were categorized as Lovers, not Family members, by our U.S. card sort participants). Indians also expressed patriotism and hope for a better future for their kids and future generations. Americans, on the other hand, often shared their emotionally-charged political opinions with their family relationships. The patriotism/political statement partition between Indian and U.S. tweets highlights a divide in what is appropriate to express regarding national affairs (\textit{patriotism} vs. \textit{politics}). %
U.S. users also often talk about how they are using social media \MRtext{as well as posting mundane updates about what is happening in their everyday lives}. They also use profanity in making highly informal complaints. In contrast, there are no similar topics for Indian users.

\subsection{Disclosures of \MRtext{Potentially Sensitive Information, Financial and Location Tweets} (RQ3)}
\label{findings_rq4}

\newtext{We next investigate whether interpersonal disclosure norms for potentially sensitive information (location and financial) differ across India and U.S. tweets.
We also sought to gain a qualitative understanding \MRtext{of how disclosure of this type of information differed.} }

\subsubsection{Financial Disclosures}
We identified tweets that disclosed financial information by training a neural network classifier to label each tweet as either "Financial" or "Not Financial." Tweets were preprocessed to remove mentions (@), hashtags (\#), and links, and to transform emojis into textual representations. They then were passed through a BERT encoder \cite{devlin2018bert} to create contextualized word representations for input into the two-layer neural network. Training data consisted of 5,200 sentences extracted from the AG's News Topic Classification Dataset \cite{Zhang2015CharacterlevelCN} combined with 600 tweets that had been hand-labeled by our team. Classifier performance was validated on a different set of 600 tweets, also hand-coded, with a final classification accuracy of 96\%. 

We found that 4.7\% of U.S. tweets and 3.7\% of India tweets contained financial information, \newtext{suggesting that U.S. tweeters are more likely than Indian tweeters to disclose financial information in the context of interpersonal relationships. A manual inspection of the tweets
by members of our research team revealed that many of them referred to the financial information of the tweet's author rather than to a person with whom the tweeter had an identifiable relationship.}
In order to identify financial tweets which primarily disclosed about \textit{a third party with whom the author has an interpersonal relationship}, we narrowed the results by removing all tweets in which a recognized relationship word did not appear within a three-word distance of a financial keyword. This resulted in 2089 US tweets and 124 India tweets. Table \ref{tab:financial_disclosures} provides a word-level analysis of financial disclosure tweets, leading to the following observations:

\begin{table}
    \centering
    \small
    \begin{tabular}{p{7em} p{5em} p{5em} p{5em} p{5em}}
         & US-all & Ind-all & US-filtered & Ind-filtered \\
         & (15975) & (1213) & (2090) & (125) \\
        \toprule
        \cellcolor[gray]{0.8}{\textbf{money}} & \cellcolor[gray]{0.8}{\textbf{11.28\%}} & \cellcolor[gray]{0.8}{\textbf{13.69\%}} & \cellcolor[gray]{0.8}{\textbf{13.54\%}} & \cellcolor[gray]{0.8}{\textbf{24.80\%}}  \\ \hline
        buy & 9.71\% & 10.39\% & 18.09\% & 20.0\% \\ \hline
        \cellcolor[gray]{0.8}{\textbf{help}} & \cellcolor[gray]{0.8}{\textbf{4.67\%}} & \cellcolor[gray]{0.8}{\textbf{7.58\%}} & \cellcolor[gray]{0.8}{\textbf{3.25\%}} & \cellcolor[gray]{0.8}{\textbf{12.0\%}} \\ \hline%
        \cellcolor[gray]{0.8}{\textbf{bought}} & \cellcolor[gray]{0.8}{\textbf{8.86\%}} & \cellcolor[gray]{0.8}{\textbf{7.25\%}} & \cellcolor[gray]{0.8}{\textbf{33.40\%}} & \cellcolor[gray]{0.8}{\textbf{26.4\%}} \\ \hline
        \cellcolor[gray]{0.8}{\textbf{pay}} & \cellcolor[gray]{0.8}{\textbf{7.19\%}} & \cellcolor[gray]{0.8}{\textbf{4.78\%}} & \cellcolor[gray]{0.8}{\textbf{12.06\%}} & \cellcolor[gray]{0.8}{\textbf{5.6\%}} \\ \hline
        \cellcolor[gray]{0.8}{\textbf{bank}} & \cellcolor[gray]{0.8}{\textbf{1.18\%}} & \cellcolor[gray]{0.8}{\textbf{3.96\%}} & \cellcolor[gray]{0.8}{\textbf{0.96\%}} & \cellcolor[gray]{0.8}{\textbf{2.4\%}} \\ \hline
        \cellcolor[gray]{0.8}{\textbf{loan}} & \cellcolor[gray]{0.8}{\textbf{0.50\%}} & \cellcolor[gray]{0.8}{\textbf{2.39\%}} & \cellcolor[gray]{0.8}{\textbf{0.81\%}} & \cellcolor[gray]{0.8}{\textbf{0.8\%}} \\\hline
        business & 2.29\% & 2.56\% & 1.00\% & 0.8\% \\ \hline
        house & 2.75\% & 2.14\% & 3.44\% & 2.4\%  \\ \hline
        give & 2.85\% & 2.39\% & 2.39\% & 3.2\% \\ \hline
        \cellcolor[gray]{0.8}{\textbf{salary}} & \cellcolor[gray]{0.8}{\textbf{0.24\%}} & \cellcolor[gray]{0.8}{\textbf{2.06\%}} & \cellcolor[gray]{0.8}{\textbf{0.19\%}} & \cellcolor[gray]{0.8}{\textbf{8.0\%}} \\ \hline
        \cellcolor[gray]{0.8}{\textbf{car}} & \cellcolor[gray]{0.8}{\textbf{2.39\%}} & \cellcolor[gray]{0.8}{\textbf{1.98\%}} & \cellcolor[gray]{0.8}{\textbf{3.06\%}} & \cellcolor[gray]{0.8}{\textbf{1.6\%}} \\ 
        \bottomrule
    \end{tabular}
        \caption{\textbf{Financial disclosure - word frequencies.} Percentages indicate the number of tweets containing the specified word. We compare the frequency of each word across both cultural groups for (1) all tweets containing financial disclosures ("-all"), whether about oneself or about a related party, and (2) tweets containing disclosures \textit{about a specific person with whom the tweeter has a personal relationship} ("-filtered"). Rows with particularly interesting discrepancies between cultural groups are shown in bold-face text marked in \colorbox[gray]{0.8}{\textbf{gray}}.}
    \label{tab:financial_disclosures}
\end{table}

\label{sec:financial_quantitative}

\begin{itemize}
\item Both India and U.S. tweets refer to "money" with approximately equal frequency. But when disclosing other people's financial information (as opposed to their own or their country's), Indian tweeters discuss money nearly twice as frequently as U.S. tweeters (24.8\% vs 13.54\%).
\item When speaking about a specific person with whom the user has a personal relationship, the word "help" occurs roughly 4 times more frequently in Indian tweets than in U.S. ones, while U.S. tweets are 1.5-2.0 times as likely to use the words "bought" and "pay".
\item With respect to interpersonal disclosures, the word "salary" is used with much higher frequency in India vs the U.S. (8.0\% vs 0.19\%). Manual inspection suggests that in India, salary is frequently referenced in the context of parental relations, such as giving one's salary to one's parents, whereas in U.S. tweets, salary is used either in complaint of one's financial status or in the context of providing for one's children. (See appendix Section \ref{sec:appendix_salary}.)
\end{itemize}

To augment the word-level analysis, we also open-coded 180 financial tweets (90 from each culture group). The open coding revealed U.S. tweets tend to disclose their financial position in a communicative way. For instance, a U.S. tweet mentioned \MRtext{\textit{``yeah me too. just finished my last payment for 3rd / last child. my babies got no school loans.''}} On the other hand, Indian tweets usually share their situation in the context of solicitations for help. Indian tweets also have a much higher number of complaints (24.4\% as compared to 13.3\%), often in the context of requesting tech support or service from institutions. This pattern did not appear in U.S. tweets.

\newtext{Key takeaways from the open coding of financial tweets are as follows:
\begin{itemize}
    \item Individual financial status is frequently shared by both cultures. However, they take different approaches. U.S. tweets tend to use a light tone of voice to share about their financial position, especially when talking about tough financial situations. For example, one U.S. tweet reads \MRtext{\textit{``I will have \$8 in my bank account after the guy comes to pop my car door open because I locked my keys in my car at the dollar store.''}} Indian tweets usually share how they are looking for help and have more instances of asking for financial assistance. For example, one Indian tweet mentioned, \MRtext{\textit{``i need a bank loan for survival my family because i am not able to walk i want open a small fast food shop for survival my family please help us''}}. 
    \item Indian tweets have a much higher number of complaints. From a sample of 90 tweets per country, 12 from the United States were hand-coded as complaints and 22 from India. U.S. complaints usually focus on a name-and-shame approach without expectation of influencing institutional behavior, while Indian tweets are often complaints requesting tech support or services from companies, banks, or other institutions. For example, one Indian tweet complained to the bank that \MRtext{\textit{``sir kindly help me out from my grandfather's account monthly a sum of money 2400 has been deducted ... so please help''}}.
    \item U.S. users tend to share about current loans and debt while Indian users frequently solicit or complain about loans from companies on their Twitter page. In our hand-coded sample, out of the 6 tweets from India referring to financial institutions, 3 were sharing complaints about wrongdoings and 3 were requesting financial assistance. For example, one Indian Twitter user reported that \MRtext{\textit{``shame on you. my niece and her office mate getting abusive call from your recovery agent for a loan her father has taken. she is not even aware of it.''}}.
    \item A common pattern of financial disclosure is talking about gifts received or purchased for loved ones and family. While most purchases are small, some Indian and U.S. tweets talk about large purchases like houses, cars, and planes. These larger purchases are more frequent in the Indian tweets than in the U.S. ones.
\end{itemize}}

\subsubsection{Location Disclosures}
To identify tweets containing location disclosures, we used spacy's named entity recognition system \cite{spacy} to identify named entities with tags that correspond to locations (e.g., 'GPE' for geopolitical entities including city and country names). 5.04\% of  U.S. tweets and 5.85\% of India tweets were found to contain location information. This would seem to suggest the Indian users were more willing than U.S. users to share location information. However, open coding on location tweets revealed that the distribution of location disclosure types varied across the two cultures. U.S. users were more willing to share their \textit{current} location. In sampling 300 tweets (150 per country), 16 (10.67\%) of the U.S. tweets share explicitly the current location of the user, compared to 0 from India.

The largest difference identified during open coding involved the association of specific past memories with locations. Memories were involved in 30.6\% of open-coded India location tweets, but only 6.6\% of U.S. location tweets. Indian tweets also share more about religious beliefs, specifically about Islam. No religious mention was found in the U.S. sample.

\subsection{Sentiment and Emotional Norms (RQ4, RQ5)}
\label{findings_rq3}
\newtext{To explore emotional norms in the context of interpersonal information disclosure, we leveraged} the \textit{ifeel2.0} multilingual benchmarking system for sentence-level sentiment analysis \cite{araujo2016ifeel}. Each tweet was classified using the ifeel2.0 system to produce sentiment scores in the range [0.0,1.0] for each of 6 emotions: Anger, Disgust, Fear, Joy, Sadness, and Surprise. \MRtext{ifeel2.0 provides multilingual support by converting non-English languages into English text (using machine translation) and then running sentiment detection methods designed for English sentences. Since our corpus is mostly monolingual (in English), we used the sentiment detection module of ifeel2.0 directly without machine translation}. When a tweet's sentiment score for a given emotion exceeded a hand-tuned threshold of 0.5, the tweet was recorded as expressing that emotion. While this method excludes the possibility of detecting multiple emotional expressions within a single tweet, our manual examination of ifeel2.0's output suggests that the 0.5 threshold corresponds well to human perceptions of emotional expression in tweets.

We begin by comparing emotional expression in tweets that passed dependency parsing (see Section \ref{sec:identifying_relevant_tweets}), and thus disclose interpersonal relationships. Results are shown in Table \ref{tab:web_tweets}. \newtext{We find that India tweets contain 2\% more emotion overall than U.S. tweets, with U.S. tweets expressing slightly more emotional content than Indian tweets in the categories of Anger, Sadness, and Surprise, while Indian tweets express slightly more Disgust, Fear, and Joy. In our study domain of tweets disclosing interpersonal relationship information, Joy was by far the most prominent emotion in both cultures, but Indian tweets expressed considerably more joy than U.S. tweets did.}

\begin{table}
    \small
    \begin{tabular}{p{7em} p{4em} p{4em} p{4em} p{4em} p{4em} p{4em} p{4em}}
    & Anger & Disgust & Fear & Joy & Sadness & Surprise & None\\
    \toprule
    India & 1.34\% & 0.44\% & 6.81\% & 42.00\% & 5.81\% & 19.02\% & 24.57\%\\
    U.S. & 2.13\% & 0.40\% & 5.90\% & 35.80\% & 7.81\% & 21.92\% & 26.10\%\\
    \bottomrule
    \end{tabular}
    \caption{Emotion expression across web-scraped tweets from January 2020 \newtext{that disclosed interpersonal relationships generally, without regard to the specific category of interpersonal relationship involved. Based on a subset of 
    695,277 U.S. tweets and 62,022 India tweets that were subsequently passed through dependency parsing to identify valid interpersonal relationship disclosures.}} 
    
    \label{tab:web_tweets}
\end{table}

\begin{figure}
\centering
\includegraphics[width=0.98\linewidth]{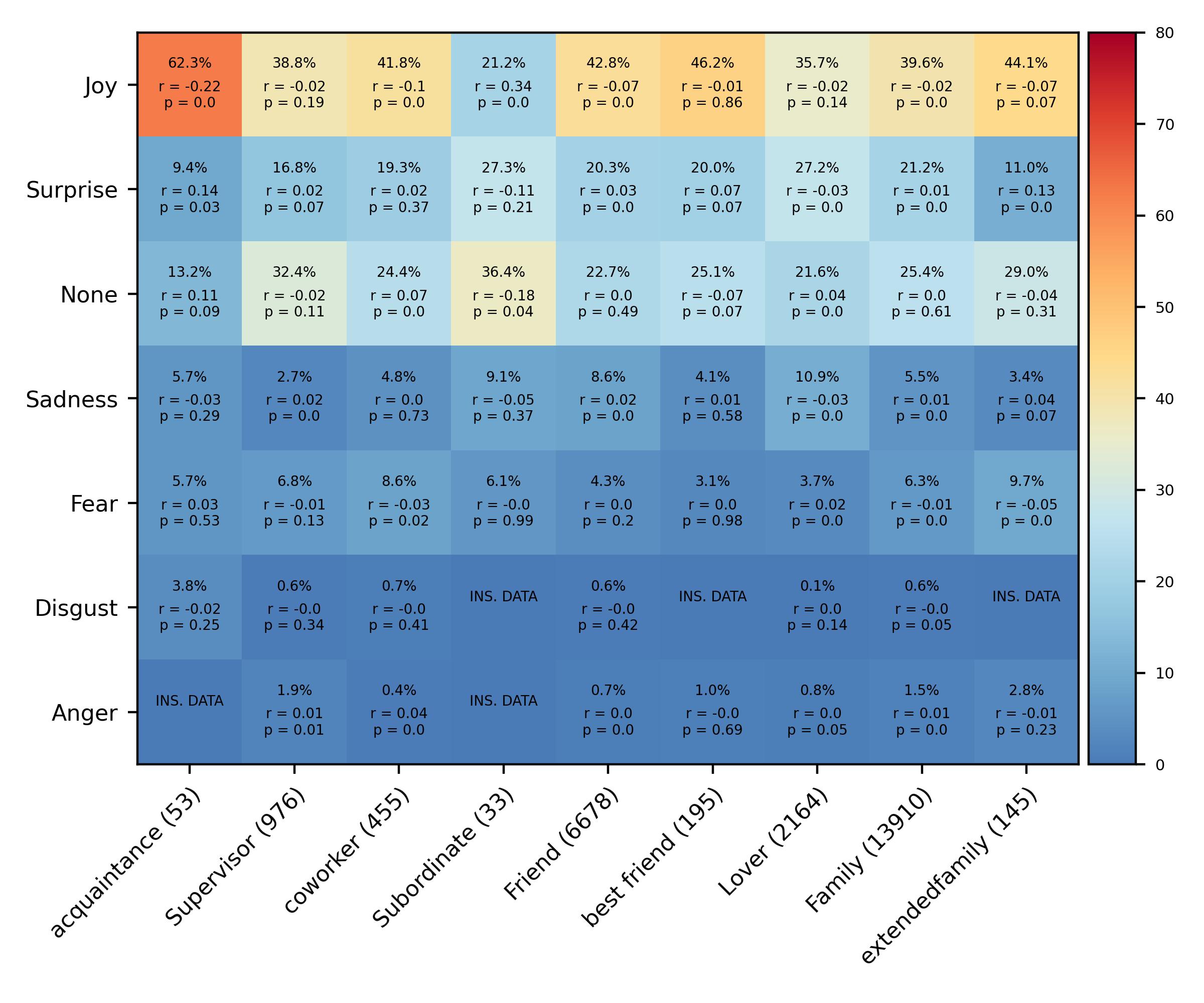}
\caption{Fine-grained emotion analysis of Indian tweets, including percentage occurrence, rank-biserial correlation between cultures, and statistical significance. \newtext{Color-coding corresponds to percentage occurrence}. "INS. DATA" stands for "Insufficient Data", \newtext{and means that at least one culture contained too few tweets for meaningful statistical analysis.}}
\label{fig:fine_grained_emo1}
\end{figure}

\begin{figure}
\centering
\includegraphics[width=0.98\linewidth]{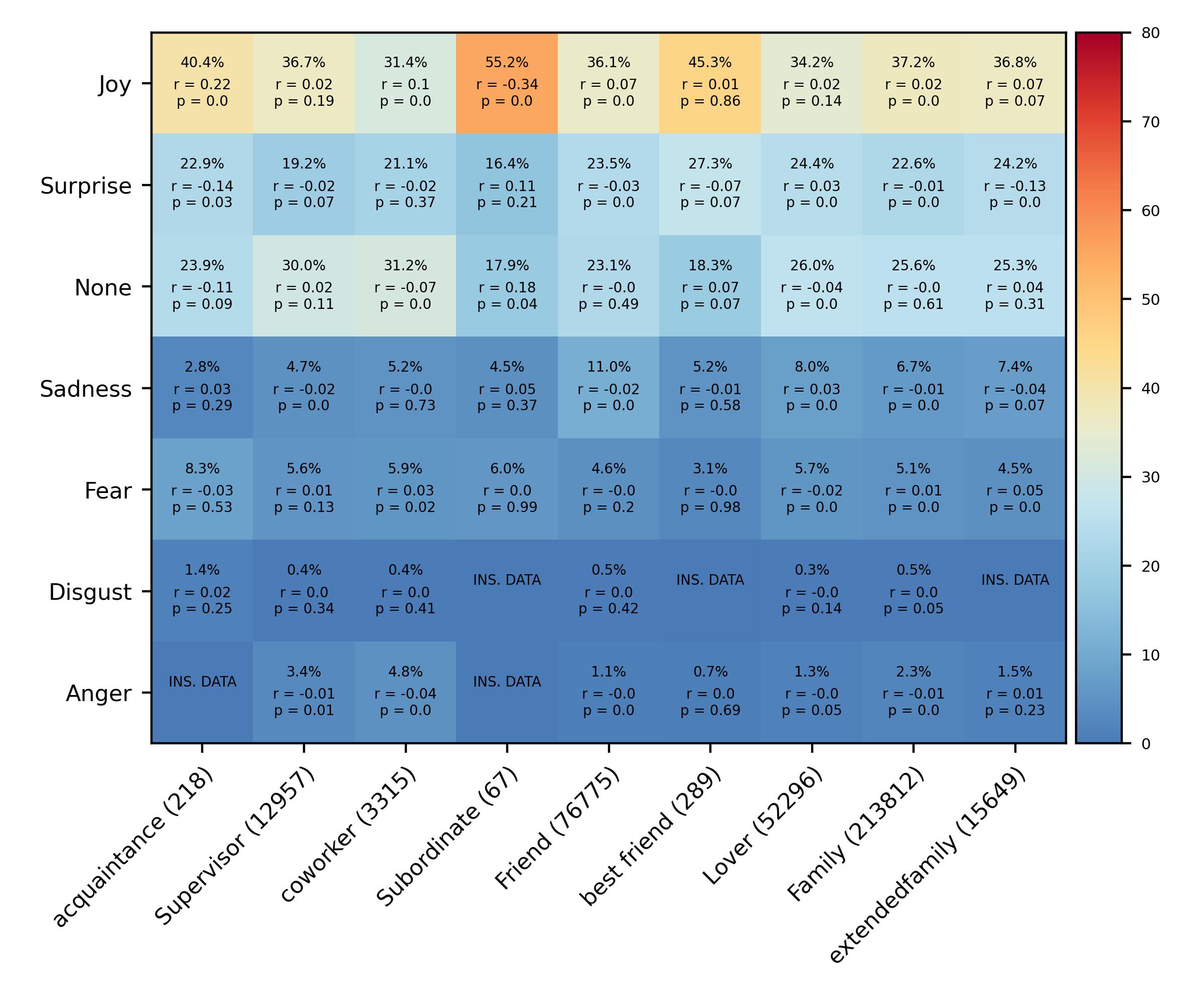}
\caption{Fine-grained emotion analysis of U.S. tweets, including percentage occurrence, rank-biserial correlation between cultures, and statistical significance. \newtext{Color-coding corresponds to percentage occurrence.} "INS. DATA" stands for "Insufficient Data", \newtext{and means that at least one culture contained too few tweets for meaningful statistical analysis.}}
\label{fig:fine_grained_emo2}
\end{figure}

We next conducted a fine grained analysis of emotional expression in the context of specific  relationships (Figs. \ref{fig:fine_grained_emo1} and \ref{fig:fine_grained_emo2}). Percentages indicate the proportion of tweets containing a given emotion, r values are the Rank Biserial Correlation for each emotion/relationship combination (with 0.0 indicating an identical distribution across both cultures), p-values show the level of statistical significance. "INS. DATA" (Insufficient Data) indicates that at least one culture group did not contain enough tweets to allow statistical analysis.

We find that when discussing specific interpersonal relationships, Indian tweets express consistently more joy than U.S. tweets across all but one of the relationship categories. \newtext{Particularly interesting category distinctions included:
}

\newtext{- \textit{Acquaintance}: Indian tweeters are 1.54 times as joyful than U.S. tweeters}

\newtext{- \textit{Co-worker}: U.S. tweeters express nearly 12 times as much anger as Indian tweeters}

\newtext{- \textit{Subordinate}: 55.2\% of U.S. tweets that refer to Subordinates express joyful emotions. This makes the subordinate category the most joyful relationship group for U.S. tweeters, and the only category for which U.S. tweeters are more joyful overall. Indian users, in contrast, express joy in only 21.2\% of tweets about Subordinates, making it the least joyful of all the Indian relationship categories.}

\newtext{- \textit{Friend}: Indian tweeters are roughly 1.19 times as joyful, while U.S. users express 1.28 times as much sadness as their counterparts from the other culture.}

\newtext{- \textit{Extended Family}: U.S. tweeters express more than twice as much surprise and nearly twice as much sadness as Indian tweeters, while Indian tweeters express 1.19 times as much joy and more than twice as much fear as their U.S. counterparts.}

\newtext{To investigate this interesting pattern inversion relating to the Subordinate relationship category, we conducted open coding on a randomly sampled selection of 200 tweets from both the India and the U.S. Subordinate categories. The additional level of joyful tweets in the U.S. Subordinate category results primarily from a high incidence of shout-outs and public praise, most frequently in conjunction with the relationship word "mentee". }

\section{Discussion of Culture-specific Norms}

\subsection{Culture-Specific Norms in Online vs. Offline Information Disclosure}

Our results point to several differences in individualistic versus collectivist cultural norms when disclosing about one's interpersonal relationships in an online platform, Twitter. While some findings were in line with earlier work on offline interpersonal disclosure examined cross-culturally, others were not. This may indicate that the norms observed in the cross-cultural studies on offline interpersonal disclosure might not always hold when communicating online, more concretely through Twitter. 

\begin{table}[]
\begin{tabular}{|l|l|l|l|}
\hline
                                                & \begin{tabular}[c]{@{}l@{}}Offline \\ (Prior work)\end{tabular} & \begin{tabular}[c]{@{}l@{}}Online\\ (Our work)\end{tabular} & Comparison \\ \hline
Communication with vertical relationships       & COL \textgreater IND                                            & IND\textgreater{}COL                                        & Disagree   \\ \hline
Positive emotion towards in-group relationships & COL \textgreater IND                                            & COL \textgreater IND                                        & Agree      \\ \hline
Negative emotion in out-group relationships     & COL \textgreater IND                                            & IND\textgreater{}COL                                        & Disagree   \\ \hline
Financial information in disclosure             & COL \textgreater IND                                            & COL \textgreater IND                                        & Agree      \\ \hline
Location information in disclosure              & COL \textgreater IND                                            & COL \textgreater IND                                        & Agree      \\ \hline
Relationship-oriented Topics                    & COL \textgreater IND                                            & COL \textgreater IND                                        & Agree      \\ \hline
\end{tabular}
\caption{\MRtext{Comparison between frequency of offline vs. online disclosure in collectivist (COL) vs. individualistic (IND) societies}}
\label{comparison}
\end{table}

\MRtext{\textit{Disclosure Frequency.   } Our results show that U.S. tweeters place greater relative emphasis on the Lover, Family, and Extended Family relationships, while Indian tweeters place greater relative emphasis on Friend, Best Friend, and Co-worker. This finding runs counter to prior work on offline interpersonal communication. Prior work has shown that collectivist cultures place higher value on their in-group relationships than out-group, thus communicating with in-group relationships more frequently in the offline settings \cite{oyserman_rethinking_2002}. Vertical relationships (i.e., parent-child, blood relatives, elders, biological family members) are more important in collectivist cultures \cite{takahashi_commonalities_2002,realo_hierarchical_1997,shkodriani_individualism_1995}. However, our findings show the opposite: horizontal relationships, such as friend, best friend, and co-worker, are more frequently mentioned on Twitter, which might indicate a difference between offline and online interpersonal disclosure on a public social media platform.}

\MRtext{\textit{Emotion Frequency.   } Our results showed that, in general, Indian users displayed more joy than U.S. users when tweeting about interpersonal information. \renewtext{This pattern is in line with prior cross-cultural interpersonal communication research in offline settings. Prior studies have shown that people in collectivist cultures tend to express more positive emotions and less negative emotion in their offline communication with in-group relationships in order to maintain in-group harmony \cite{triandis_individualism_1988,kitayama_cultural_2006,matsumoto_mapping_2008,takahashi_commonalities_2002}. Since most interpersonal relationships examined in our paper were in-group relationships, such as family, friend, and co-worker, we confirm such pattern of positive emotions in the online interpersonal communication.}}

\newtext{However, breaking down the emotion findings by relationship type revealed some big differences. For the Subordinate relationship category, U.S. users showed unusually high amounts of joy while Indian users showed unusually low amounts. Further qualitative analysis suggested that there may be a U.S.-specific norm of praising one's mentee on Twitter. \renewtext{Supervisor-subordinate communication has been less studied in prior cross-cultural communication research in the offline setting. One possible reason behind this finding is that \renewtext{the supervisor-subordinate \MRtext{relationship context} in a collectivist culture such as} India emphasizes hierarchical status and social order. Individuals in higher status, such as supervisors, are more restrictive towards their subordinates \cite{vogel2015cross}, less participatory, more authoritarian, and more directive in collectivist cultures \cite{dickson2003research}.} This result suggests a specific online practice that has evolved on Twitter and become an acceptable norm in the U.S., but not in India. New communication channels such as Twitter can enable such expressions which may not have had an outlet offline. \renewtext{Consequently, our study} reinforces the importance of studying emerging practices and new norms of disclosure that can rise with novel communication mechanisms. }

\newtext{Additionally, U.S. tweeters expressed 12 times as much anger related to co-workers than Indian tweeters. \renewtext{This is in line with the cultural norm of interpersonal disclosure in the offline settings that speakers from collectivist cultures tend to use positive expression towards in-group relationships such as co-workers \cite{matsumoto_mapping_2008}. Other findings ran counter to this cultural norm, but to a lesser extent. For instance, Indian tweets about acquaintances were one and a half times as joyful as U.S. tweets, which disagrees with prior work that collectivists create greater distance from out-groups, such as acquaintances and strangers, by using more negative emotional expression \cite{matsumoto_mapping_2008}.} Again, this may be an evolving practice that is facilitated by this new communication medium. From a methodological viewpoint, our findings suggest that when examining emotional interpersonal disclosure norms, a fine-grained \renewtext{and more relationship-specific contextual} analysis across multiple relationship categories will uncover more meaningful insight than aggregate statistics across all tweets and relationship types.}

\MRtext{\textit{Information Disclosure Content and Frequency.}   }  \MRtext{The frequency of disclosing sensitive information does not match} prior cross-cultural privacy literature. \renewtext{Prior work on offline cross-cultural disclosure} has shown that people in individualistic cultures are more concerned with and less likely to disclose sensitive personal information on social media than those in collectivist cultures \cite{chen_differences_1995,cho_qualitative_2013,posey_proposing_2010,tsoi_privacy_2011}. Our quantitative results suggested that this is not supported. \MRtext{However, upon deeper analysis, we find that the \textit{content} of financial disclosures differed between the two cultures in a way that aligns with the theory. }\renewtext{Indian tweets were much more likely to discuss other people's finances, whereas U.S. tweets were more likely to be complaining about one's own salary. This \renewtext{result} aligns with \renewtext{general} collectivist versus individualistic attitudes. \MRtext{In India,} the much higher occurrence of the word ``help'' suggests that interpersonal financial disclosures in India may have a greater emphasis on collectivist principles such as giving to and receiving help from other members of the community. Indeed, Indian tweets were much more likely to discuss other people's finances. Similarly, the word ``salary'' referred more to giving money to one's parents, whereas in the U.S., it was more likely to be complaining about one's own salary. In fact, the Indian tweets tended to be action-oriented to help others or seek support, while U.S. tweets were more often describing their own financial situation. Overall, we found that the quantity of discussion about financial topics did not necessarily reflect social norms so much as does the content itself. This finding suggests that the purpose of financial disclosures align with cultural values.}

The quantitative analysis of location data showed that Indian tweets would reveal more location data, which agrees with prior privacy research that individualists are more concerned with and are less likely to disclose sensitive personal information on social media \cite{chen_differences_1995,cho_qualitative_2013,posey_proposing_2010,tsoi_privacy_2011}. \MRtext{They usually have a wider variety of online social networks on social media \cite{kim2011cultural} and thus are protective about private information. However, qualitative analysis revealed that U.S. tweets more often contain information about the user's current location, while Indian tweets more often was a memory of a past location. Ironically, when current location information is shared it is potentially more sensitive since it is one's current location which can be used to locate the individual.} Sharing location data about \MRtext{past memories} might help users in a collectivist culture maintain their relationships over \MRtext{Twitter}. Thus, it's less clear whether U.S. tweets really are more privacy sensitive. These findings have implications for those studying online cross-cultural disclosures of sensitive data. Our work makes it clear that researchers should look into the content and not just rely on the numbers to understand disclosure attitudes and behaviors of their subjects.

\newtext{Finally, the topical analysis of tweets revealed interesting trends that align with expected cultural attitudes. For instance, the tendency of U.S. users to complain about family members, use profanity while engaging in informal complaints, talk about self-promotion through social media activities, praise their spouse, and discuss their political opinions, align with individualistic values. In contrast, Indian tweets emphasized expressing love about their in-group relations, talking about kids and posterity and patriotism. Conspicuously lacking was complaints about family or narratives around social media self-promotion. This aligns with a collectivist approach of promoting harmony for in-group relations, connection with a broader in-group, and expressing positive sentiments. These different norms may be used as a guide for appropriate content \MRtext{(and as a starting point to implicitly identify} inappropriate content) in each culture. 
}

\newtext{Given all of these different cultural norms that \MRtext{ we have observed in} Twitter disclosures, we emphasize to researchers and designers that a violation of such \renewtext{disclosure} norms could lead to privacy violations. Our research establishes a baseline \renewtext{(using both quantitative and qualitative analysis)} for normative behavior when disclosing interpersonal information on Twitter. This can be used to anticipate, identify, or design against potential and actual privacy breaches. We next turn to our recommendations for doing so.}

\MRtext{\subsection{Contextual Information Norms in Interpersonal Disclosure}}

\MRtext{Our findings indicate that the contextual information norms around information type and actors in interpersonal disclosure differ between U.S. and India. According to the contextual integrity framework \cite{nissenbaum-2010-ci-book}, information flows are governed under the contextual information norms in a specific context. Context can be conceptualized using three parameters: actors (subject, sender, recipient), attributes (types of information), and transmission principles (constraints under which information flows). When these parameters change, the context changes, and the contextual information norms change. Information flow that is appropriate in one context may become inappropriate in another context. In our study, the major contextual factors examined are information types (i.e., emotions and topics in the tweets about interpersonal disclosure) and information subjects (i.e., the relationship types mentioned in the tweets). Indian users exhibited more joy and disclosed more financial/location topics than U.S. users when tweeting  interpersonal information.  Additionally, U.S. tweeters place greater emphasis on the Lover, Family, and Extended Family, while Indian tweeters place greater relative emphasis on Friend, Best Friend, and Co-workers. This suggests that the norms around information type and subject vary between U.S. and India. }

\MRtext{The effect of contextual factors on users' privacy attitudes and behaviors have been extensively examined in prior work. For example, users are more likely to disclose information to close ties than to distant ties \cite{consolvo2005location}. Personal information, such as activity data \cite{bilogrevic2013adaptive} locations (i.e., bank, hotel, etc.) \cite{dong2015predicting}, and inappropriate content (i.e., drinking alcohol) \cite{fogues2017sharing} are less shared on social media. However, most of these works investigated contextual information norms in a single country, mostly western countries. And the contextual information norms examined in prior work are mostly about individual information sharing, rather than disclosure about interpersonal information. Our work fills these gaps by showing that the contextual information norms in interpersonal disclosure vary between different countries. This suggests that the contextual information norms identified in previous work may not directly apply to other countries like India. Future cross-country privacy research should look at different sets of contextual factors in different cultures which shape interpersonal disclosures. }

\MRtext{Additionally, our work contributes to the multi-party privacy literature by examining how users disclose about interpersonal relationships in textual information sharing. Prior work on multi-party privacy mostly focused on sharing images on social media, namely group photo sharing. Researchers have found that users are concerned when their unwanted images, such as drinking and inappropriate dressing, are shared to known social circles like family members and employers, as well as unknown audiences. Compared to these prior studies, our study focuses on the multi-party privacy in textual information sharing. Our findings reveal that while textual information might not be as identifiable as image information, it still reveals co-owner's personal information to a public audience. Users tend to share their relationships with others, certain emotions to a larger extent, and specific topics such as finance and location. Thus, it is important to design features to help make interpersonal disclosures more visible to co-owners, akin to features such as tagging and linking for photo sharing. We suggest future work investigates this topic, including co-owners' attitudes and preferences regarding textual interpersonal disclosure in order to inform the design of effective privacy management features.}

\MRtext{\subsection{Contextualizing Cultural Differences in Interpersonal Disclosure norms from Indian Twitter users}}

\MRtext{We pause to expand on how our sample of Indian users compares with prior Indian studies and what it means for our results. Moreover, we reflect on whether the collectivist-individualistic lens is
suitable for detecting interpersonal disclosure norm differences between India and U.S. In our search, we found that Indian Twitter data is often used to gauge the Indian population's opinions regarding important issues such as the election or COVID~\cite{venigalla-2020-covid-twitter-india, PAL201797, IJoC6705}. Moreover, earlier work has quantified general attitudes of Indian users through the lens of individualism and collectivism. While  research shows that today's Indian youth may appear to endorse western (and more individualistic) values, the family traditions, group values, and national traditions still play an extremely important role in their decision making process~\cite{khare-2011-brand-india}. This phenomena might be facilitated by the fact that in India most homes contain older relatives (grand/parents, uncles, and general 'family community') who shape these users.  In fact, Ramamoorthy et al. identified that Indian employees from even high-technology sectors indeed show more collectivist traits than their Irish (more individualistic) counterparts~\cite{RAMAMOORTHY2007187}. In Ramamoorthy et al.'s study, Indian employees showed higher levels of normative and affective commitment. Our study sample may be similar and thus have similar tendencies since the Twitter users considered in this work (by design) speak English and thus are likely to be younger, more urban, and tech savvy, similar to the population studied by Ramamoorthy et al. Indeed, our findings reflect this similarity in that U.S. tweeters expressed nearly twelve times as much anger as Indian tweeters about their co-workers. Thus, this prior work suggests that our sample can be representative of more collectivist values for the purposes of our research.}

\MRtext{Finally, earlier work suggests that societal developments in India have shifted social values for the general Indian populace. Jha et al. note that at higher urbanization levels, collectivism level is lower~\cite{jha-2011-north-india}. Our Indian Twitter users are potentially more urban and have a lower level of collectivism than the general Indian populace. This makes the differences we found between India and U.S. using an individualism-collectivism lens a conservative estimate. For the general Indian populace, the collectivism level would be higher and thus our results likely provide only a lower bound on the inter-cultural differences in interpersonal disclosure between India and the U.S.}

\section{Implications}
\label{implications}

\newtext{The design implications of our results overwhelmingly point towards considering collectivist-friendly features when designing social media. Given the different norms around emotional and topical disclosure, it would be very easy to offend or come across as insensitive to someone in another culture. For example, it would be vital to make sure that a "mad" or "dislike" emoji won't be inadvertently interpreted as referring to one's in-group relations. Similarly, while complaining about family or politics might be acceptable in an individualistic society, this may be considered unacceptable or extremely rude in a collectivist context. This might be mitigated by designing an interface that provides mutual education for users from either culture. Design efforts are needed to share norms around what is acceptable in other cultures as it pertains to a given tweet they are viewing or one they are about to post. Accounting for cultural context could help users dispel misunderstandings and avoid unnecessary conflict.}

\newtext{Methodologically, our use of data analysis combining computational classifiers and human-coding / validation lays groundwork for future research directions in computer-mediated communication. Critically, when applying neural classifiers to data we relied not only on the classifier's training loss as an estimation of its correctness, but also on human-coding and validation of the classifier's outputs. This approach lends confidence that the classifier's results are not only accurate to its training data, but are also applicable in the specific domain of research to which it is applied. Using this method, we were able to quantitatively explore qualitative characteristics (i.e., sentiment, emotion) across a large-scale dataset of more than 400K tweets, opening the doors to greater insights regarding the study domain.}

\newtext{An especially critical insight arising from our work is the way collectivist verses individualist social norms influence categorical groupings of interpersonal relationships (e.g., whether Grandma is family or extended family). This suggests opportunities for improved social media design. For example, social media platforms might recommend potential new contacts based on culturally-relevant relationship groupings. Our approach of creating culture-specific relationship taxonomies could be applied when designing for other cultures to create these relationship groupings in a culturally-sensitive way.}

\newtext{In particular, we encourage designers coming from individualist societies to consciously consider the collectivist cultural norms that may be relevant for many of their users rather than relying on default assumptions that may be deeply rooted in individualistic norms. A particularly fertile area for such introspection includes the design of algorithms driving the display of specific posts or meta-information in users' input streams. For example, posts expressing anger about coworkers may not be surprising to U.S. Twitter users, but may perhaps be culturally jarring to Indian users.}

\section{Conclusion}
\newtext{This study focuses on the differences between Indian and U.S. interpersonal disclosure norms on Twitter. Using our saturated, culturally-sensitive taxonomy of online personal relationships, we found statistically significant differences in interpersonal information disclosure patterns via topical analysis, emotion- and sentiment-based analysis, and word-level analyses of potentially sensitive information disclosures. These results have implications for social media interface design, in particular with regard to the automated grouping of contacts and the display algorithms driving which posts are seen by which users. We hope that our results will help drive the development of culturally-sensitive user interfaces.}

\newtext{Additional, we encourage mixed-method analyses in cross-cultural media studies which combine manual qualitative examination of tweets with large-scale automated analysis of tweet content. This allows scholars to examine interpersonal disclosure norms at both a macro and micro scale and facilitates fine-grained examination of the data on a relationship-by-relationship basis, something which would be difficult to accomplish using hand-coding methods alone. We also present a simple but effective dependency-parsing based method of identifying tweets that disclose information about third parties with whom the tweeter has a specific interpersonal relationship, which we hope other researchers will leverage for large-scale analysis of online disclosure patterns.}

\newtext{Throughout our study, we have seen indications that individualistic and collectivist attitudes are a key factor in shaping online interpersonal disclosure norms. Since misaligned communication norms can lead to conflict and negatively impact one's relationships across cultures, we urge social media designers to go beyond the individualistic norms that are currently prominent in social media and to consider collectivist norms and values. By doing so, they can improve cross-cultural online communication and steer users away from potential misunderstandings rooted in normative differences across cultures.}

\bibliographystyle{ACM-Reference-Format}
\bibliography{cultural-disclosure}

\section{Appendix}
\appendix

\section{NOTES ON METHODOLOGICAL PROCESS}
\subsection{Methodological Process}
\newtext{A key component of our research method is the integration of computational classifiers with human-coding / validation in order to quantitatively examine fundamentally qualitative properties of the text. We combine this method with traditional open-coding methods, thus providing a comprehensive window of observation into key research areas.}

Because this methodological paradigm is not yet widespread, we provide a brief overview of our basis for confidence in the method's results prior to reporting the results themselves. In this study, we apply three neural classifiers identifying the magnitude of emotional, financial, and locational content within each tweet, as well as an LDA algorithm \cite{blei-2003-lda} for automated identification of tweet topics. Each classifier was evaluated not only based on its training accuracy, which reflects primarily how well-suited it is toward its training data, but was also hand-evaluated by a member of our team to confirm that it generalized well to our domain of interest (i.e. tweets from multiple cultures). Only classifiers that performed well under this analysis were used in our study. 

\newtext{We complement the numerical analysis provided by the classifiers and LDA algorithm with fine-grained open coding exercises to identify not only how frequently specific qualities appear in the dataset, but also what qualitative properties manifest within each subset of the data. Our goal with this method is to explore patterns of disclosure at a more fine-grained level than would otherwise be possible in order to obtain deeper insight into the phenomena involved.
}
\chngnum{7}

\section{Demography of Card-sort participants} \label{sec:demog}

\begin{figure}[!h]
\centering
\includegraphics[width=0.6\linewidth]{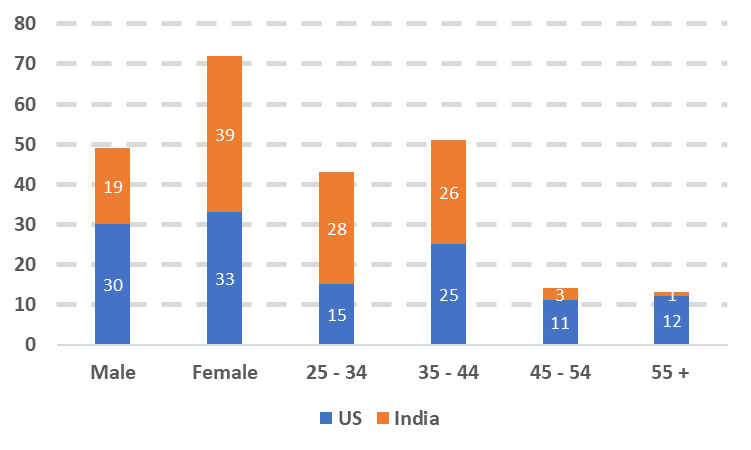}
\caption{Demographic representations of card sort participants.}
\label{fig:demogcardSort}
\end{figure}

We present the key demography of our 121 card-sort participants in Figure~\ref{fig:demogcardSort}. 

\section{CULTURE-OBLIVIOUS TAXONOMY OF INTERPERSONAL RELATIONSHIP WORDS}
\label{sec:appendix_lexicon}

\begin{table}[tbh]
    \centering
    \footnotesize
    \begin{tabular}{p{10em} p{12em}}
    Initial relationships & Final relationships\\
    \toprule
    Co-worker (peer) & Co-worker \\
    Acquaintance & Extended Family \\
    Best friend & Family \\
    Extended family & Friend \\
    Family & Lover\\
    Friend & Subordinate\\
    Lover & Supervisor\\
    Stranger & Blood Related\\
    Subordinate & Combined Relationships\\
    Supervisor & Grand Relationships\\
    & Great Grand Relationships \\
    & Spouse Relationships \\
    & Step Relationships\\
    \bottomrule
    \end{tabular}
   \caption{Left column: Initial relationship groupings based on similarity analysis. Right column: Final groupings based on more fine-grained analysis. These were used for quantitative analysis performed in Section~\ref{sec:results}.}   \label{tab:relationship_groupings}
\end{table}

\begin{table}[h]
    \footnotesize
    \centering
    \begin{tabular}{p{4em} p{4em} p{8em} p{6em} p{8em} p{4em}}
    Family & & Extended Family & Blood Related & Spouse Relationships & Lover \\
    \toprule
    baby & mum & distant relatives & aunt & wife & boyfriend\\
    brother & mummy & extended family & auntie & hubby & darling\\
    child & offspring & & aunty & husband & fiancee\\
    children & pa & & cousin & spouse & girlfriend\\
    dad & papa & & nephew & better half & honey\\
    daddy & parent & & niece & bride & hun\\
    daughter & sib & & uncle & bridegroom & love\\
    elder brother & sibling & & & & lover\\
    elder sister & siblings & & & & sweetheart\\
    fam & sis & & & & sweety\\
    family & sister & & & & true love\\
    father & son & & & & love of \\
    ma & twin brother & & & &...my life\\
    mama & twin sister & & & &\\
    mom & younger brother & & & &\\
    mommy & younger sister & & & &\\
    mother & & & &\\
    \bottomrule
    \end{tabular}
    \caption{Lexicon of relationship words identified in Section \ref{sec:lexicon} of the main paper. (Part 1/3)}\label{tab:taxonomy_full_1}
\end{table}

\begin{table}[h]
    \footnotesize
    \centering   
    \begin{tabular}{p{9em} p{9em} p{10em} p{10em}}
    Step Relationships & Grand Relationships & Great Grand Relationships & Combined Relationships\\
     \toprule
    adoptive father & grampa & great aunt & kin\\
    adoptive mother & grandchild & great granddaughter & kindred\\
    half brother & grandchildren & great grandfather & kinfolk\\
    half sister & granddaughter & great grandmother & kinship\\
    step brother & grandfather & great grandparents & kinsperson\\
    step daughter & grandma & great grandson & kith\\
    step father & grandmother & great uncle & relatives\\
    step mother & grandpa & & infant\\
    step sis & grandparent & & descendant\\
    step sister & grandparents & & kid\\ 
    stepbro & grandson & & toddler\\
    stepchild & granny & & maternal\\
    stepchildren & & & fraternal\\
    stepdad & & & paternal\\
    stepmom & & & \\
    stepparent & & & \\
    stepson & & & \\
    \bottomrule
    \end{tabular}
    \caption{Lexicon of relationship words identified in Section \ref{sec:lexicon} of the main paper. (Part 2/3)}\label{tab:taxonomy_full_2}
\end{table}
    
\begin{table}[h]
    \footnotesize
    \centering
    \begin{tabular}{p{7em} p{7em} p{7em} p{7em}}
    Supervisor & Coworker & Subordinate & Friend\\
    \toprule
    boss & colleague & mentee & bro\\
    manager & coworker & subordinate & bosom buddy\\
    master & peer & & sidekick\\
    mentor & & & bestie\\
    supervisor & & & \\
    teacher & & & \\
    \bottomrule
    \end{tabular}
   \caption{Lexicon of relationship words identified in Section \ref{sec:lexicon} of the main paper. (Part 3/3)}\label{tab:taxonomy_full_3}
\end{table}

In addition to the culture-specific relationship taxonomy identified during the card sort study, we also developed a set of saturated, culture-oblivious word lists, meaning that each relationship category contains as many words as can with confidence be included in that category across both relevant cultures, but no words whose inclusion status is doubtful or disputed. This creates confidence that comparisons between categories and across cultures are reliable and meaningful. 

The card sort task had seeded the task with ten initial relationship categories  (Table~\ref{tab:relationship_groupings} - left column) for participants to categorize similar relationship keywords. %
Beginning with this set of core categories, We %
excluded categories that had no relationship words in them, were repetitions of existing categories, or typos. We then performed a similarity analysis to identify which relationship words had been placed in a group. Namely, if greater than 80\% of participants placed a particular word within a category (e.g., `Mom' placed in `Family') that word was considered associated with that relationship group. Out of the total 178 relationship words there were 81 words in U.S. and 112 words in India that had a clear majority grouping. These words were therefore considered associated with the relationship groups they had been placed in.

For words that did not have a clear majority we applied the following rules to identify in which buckets they should be placed:

\begin{enumerate}
    \item A word sorted by 80\% of participants into two categories was assigned to the dominant one, as long as there was > 50\% agreement on that category.
    \item Other words were removed since they were spread across three categories and did not have a dominant bucket.
    \item Relationship groups with two or less relationship words were removed.
\end{enumerate}
	
Based on these rules we removed Stranger and Acquaintance categories from further analysis. We combined Best Friend with Friends since it is a specialized case of Friends and had few keywords. We then examined the remaining 8 buckets and found that while a number of relationship words placed within Family and Extended Family buckets had a majority agreement (>80\%), the words themselves differed between India and U.S. For instance, most U.S. participants had placed grand relationships (grandfather, grandmother, etc.) in the Extended Family bucket while participants from India had placed them in the Family bucket. There were similar differences for step- and adoptive relationships. We therefore only kept words in the categories if they were consistently categorized across both cultures. We then moved the other words into new categories to allow more nuanced and fine-grained categorization and analysis (i.e., Blood related, Grand Relations, Great grand Relations, and Step Relations). 

We found a similar pattern between Family and Lover buckets. For instance, Indian participants were more likely to place Spouse within the Family bucket while U.S. participants placed them in the Lover bucket. We therefore created a separate category for Spousal relationships. The taxonomy of relationship words also contained words that represent an individual's relationship with their community and larger networks (e.g., kin, relatives, descendent). While these words were mostly categorized into Extended Family or Friend categories, the agreement score for the dominant category was between 50-60\%. We therefore created a separate category Combined Relationship for these words. The objective in doing so was to allow these words to be analyzed separately from the undisputed words, in order to determine whether follow the same normative patterns. 

Our complete culture-sensitive saturated taxonomy is presented in Table~\ref{tab:taxonomy_full_1}, Table~\ref{tab:taxonomy_full_2} and Table~\ref{tab:taxonomy_full_3}.

\subsection{TWEET BREAKDOWN BY CATEGORY USING CULTURE-AGNOSTIC TAXONOMY}
\label{sec:appendix_tweet_breakdown}

\begin{figure}
\centering
\includegraphics[width=0.8\linewidth]{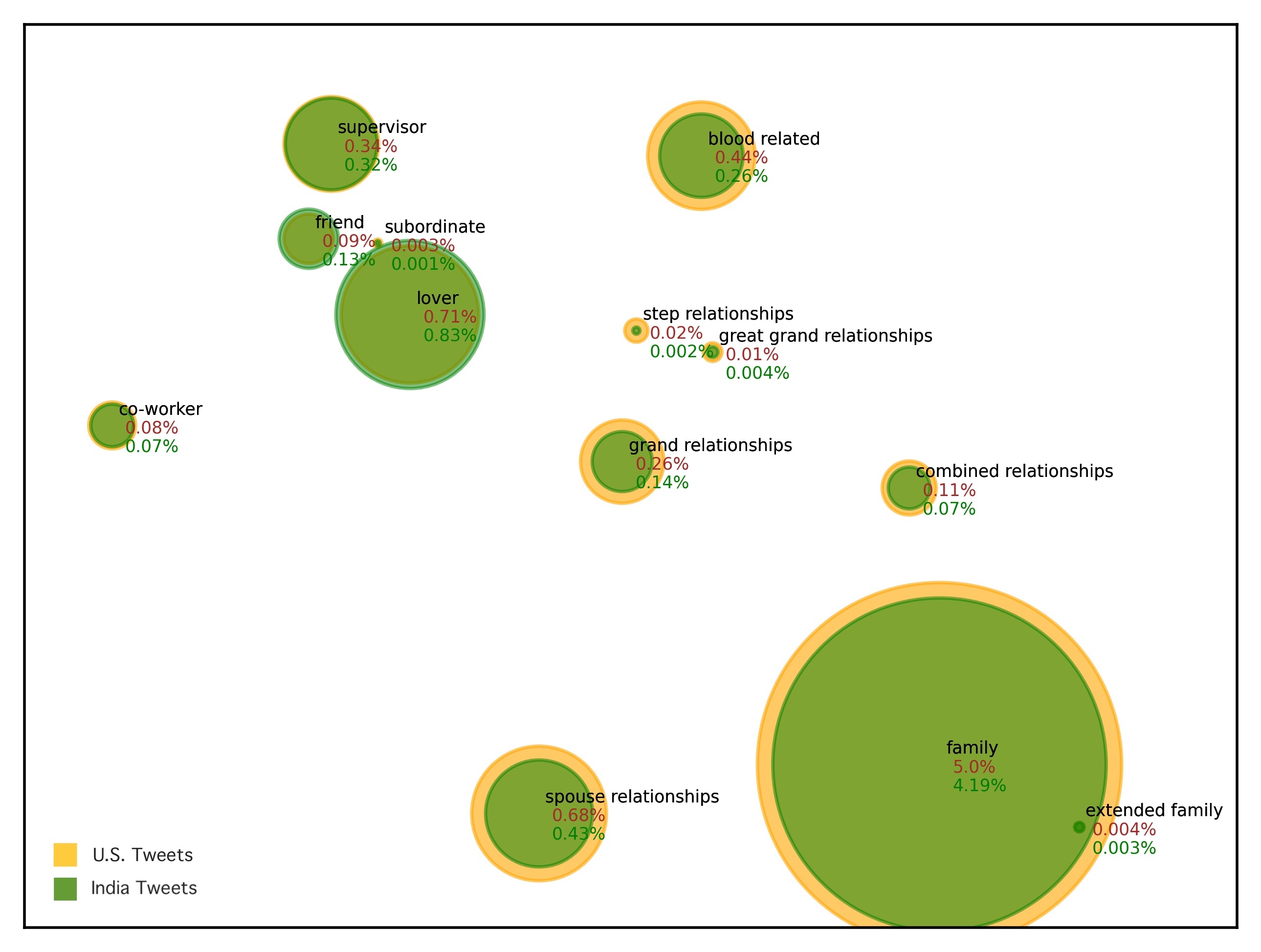}
\caption{Percentage-wise breakdown of 306,388 U.S. tweets and 23,080 India tweets using the culture-oblivious word lists. US tweets are shown in yellow (top \%), India tweets in green (bottom \%). \renewtext{The x-axis indicates each category's level of family/social relatedness, as measured using GloVe representations. The y-axis was chosen to maximize visibility between categories.}}
\label{fig:dataset_all_saturated}
\end{figure}

 A word-level (as opposed to relationship-level) analysis of relationship tweets using the identified saturated word list (see Appendix, Figure \ref{fig:dataset_all_saturated}) reveals additional interesting patterns. For example, U.S. tweeters spend more of their relationship-based tweets referencing vertical in-group relationships than Indian tweeters, a result that contradicts H2b. In contrast, horizontal in-group relationships appear to be equally represented in both cultures, with Indian tweeters being more likely to reference lover and friend relationships (as defined by the saturated word lists) and U.S. tweeters being more likely to reference spouse, family, and blood related relationships.

 This data shows that U.S. users more frequently make online disclosures of vertical relationships (Blood Related, Grand Relationships, Great Grand Relationships) while Indian and U.S. users show disparate behavior on horizontal relationships. For example, Indian tweets discuss Friend relationships more frequently, but U.S. tweets are more likely to refer to Spouse Relationships.

\subsection{EMOTION ANALYSIS OF TWEETS USING THE CULTURE-OBLIVIOUS TAXONOMY}

\begin{figure}[!h]
\centering
\includegraphics[width=0.8\linewidth]{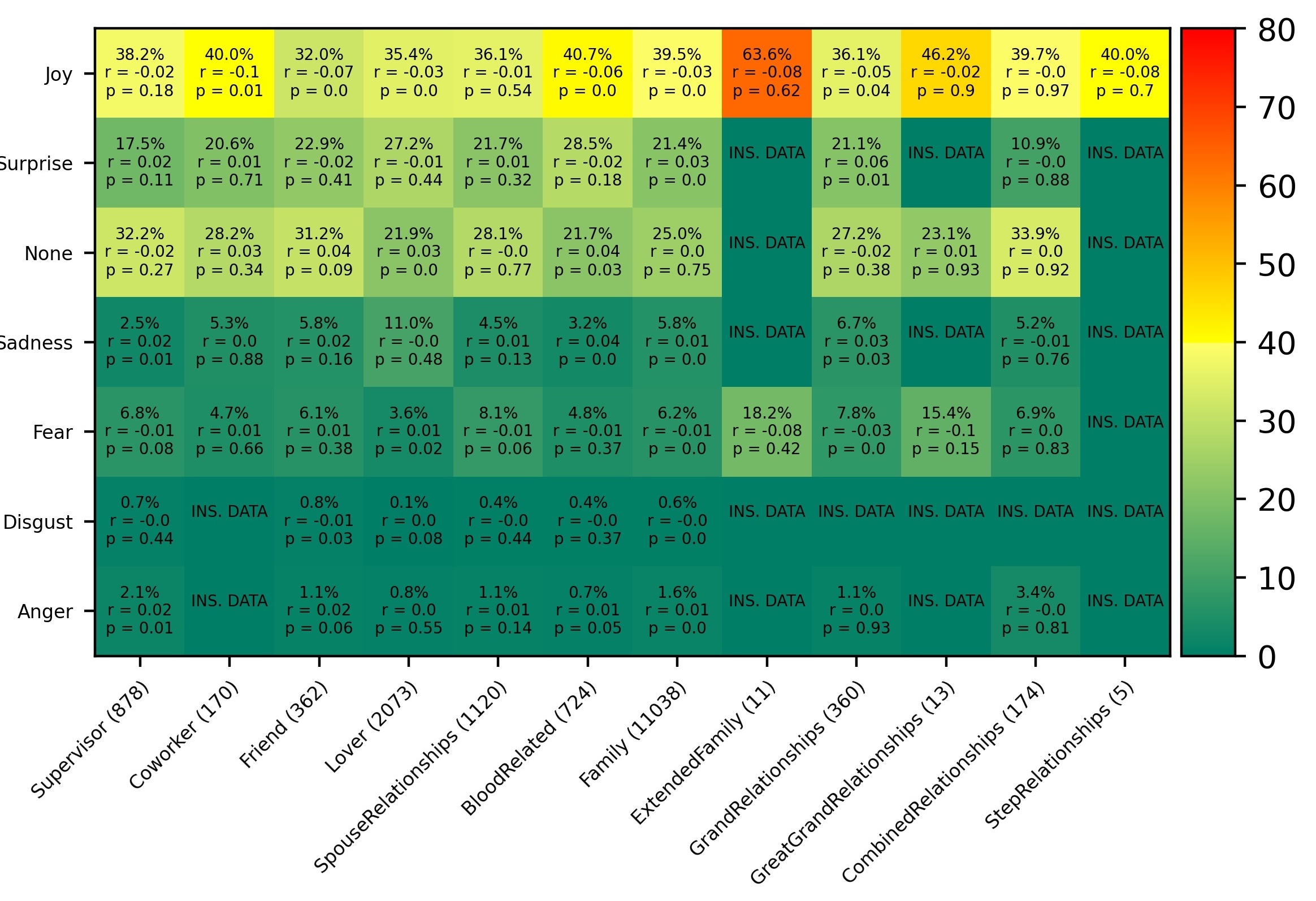}
\includegraphics[width=0.8\linewidth]{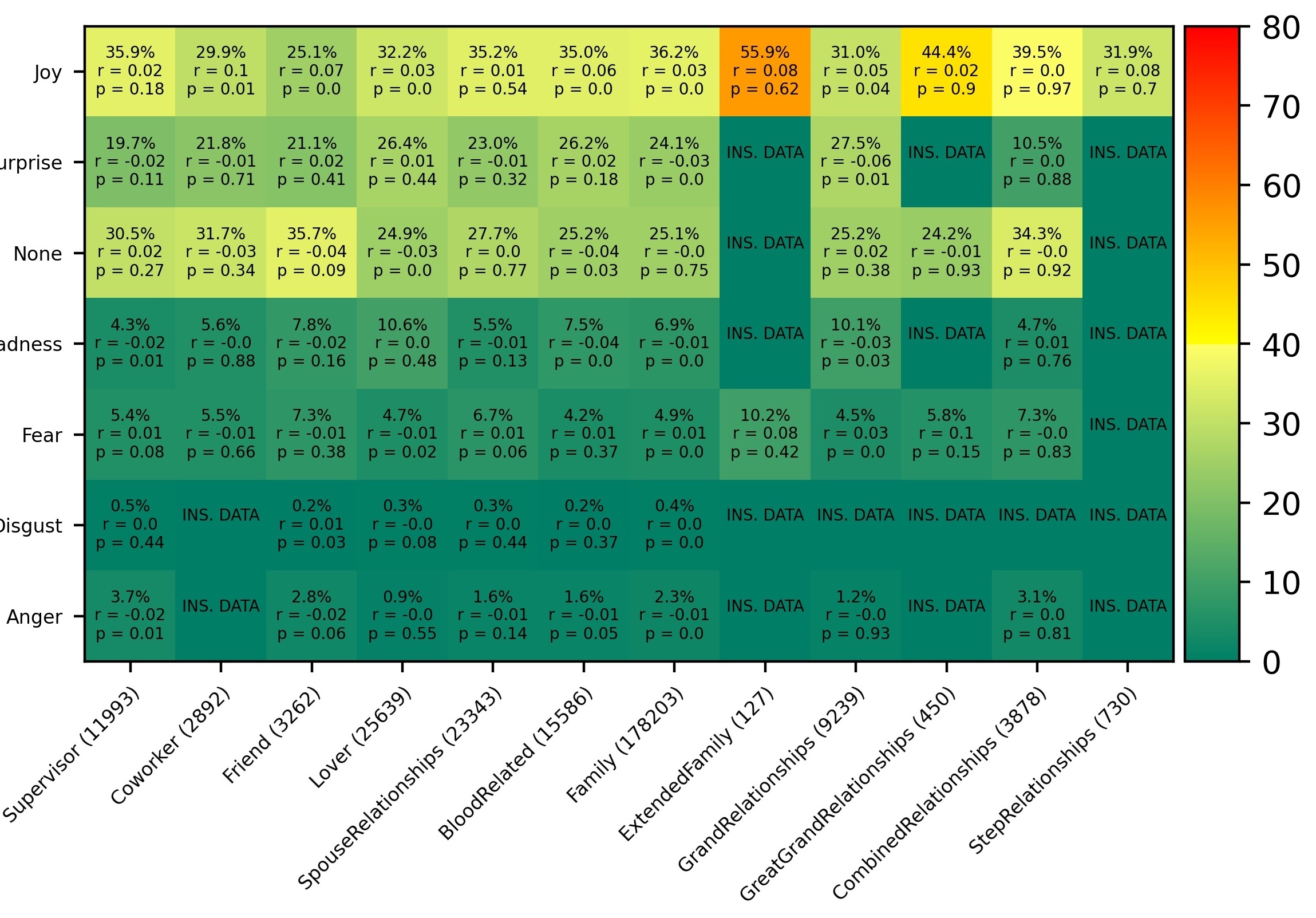}
\caption{Fine-grained emotion analysis of Indian tweets (top) and U.S. tweets (bottom), including percentage occurrence, rank-biserial correlation between cultures, and statistical significance. Color-coding corresponds to percentage occurrance. "INS. DATA" stands for "Insufficient Data", and means that at least one culture contained too few tweets for meaningful statistical analysis. Plots were generated using the culture-oblivious word lists.}
\label{fig:fine_grained_emo_saturated}
\end{figure}

Emotional breakdown by category for the culture-oblivious relationship taxonomy is given in Figure~\ref{fig:fine_grained_emo_saturated}. Many of the culture-specific patterns found in Figure~\ref{fig:fine_grained_emo1} and Figure~\ref{fig:fine_grained_emo2} (e.g., unusually positive sentiment for the US ``subordinate'' category; unusually positive sentiment for the Ind ``acquaintance'' category) are not evident in the saturated relationship groups, suggesting that key differences in which individuals are classified as part of a relationship are a driving force in the emotional landscape of various relationship groupings.

Of particular note is the uncommonly high amount of joy across both cultural groups for words in the ``Extended Family'' and ``Great Grand Relationships'' categories. 
The coworker and step relationship categories are also interesting, as both of them have a greater relative amount of joyful tweets in India, but proportionately lower amounts of joy in the US.

\section{SENTIMENT-LEVEL BREAKDOWN OF RELATIONSHIP TWEETS}\chngnum{18}
\label{sec:appendix_sentiment}

\begin{figure}
\centering
\includegraphics[width=1\linewidth]{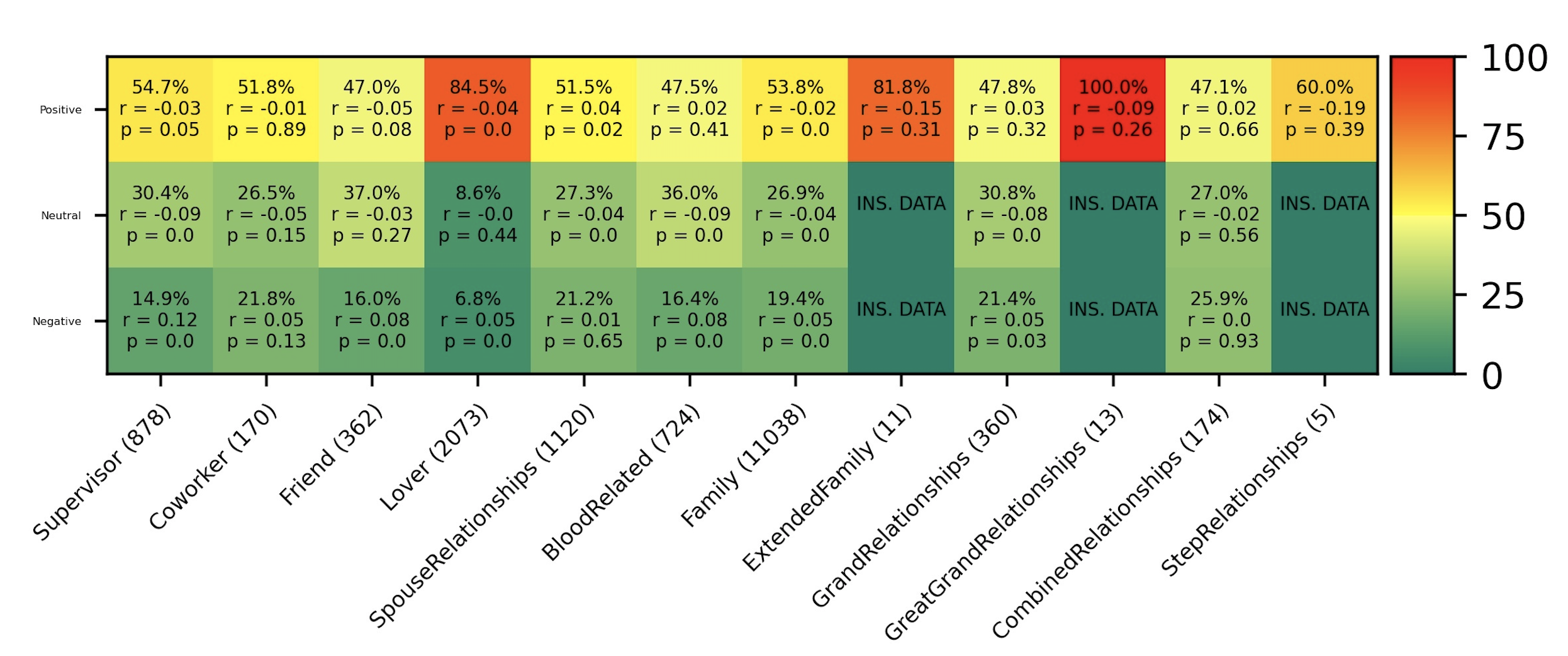}
\includegraphics[width=1\linewidth]{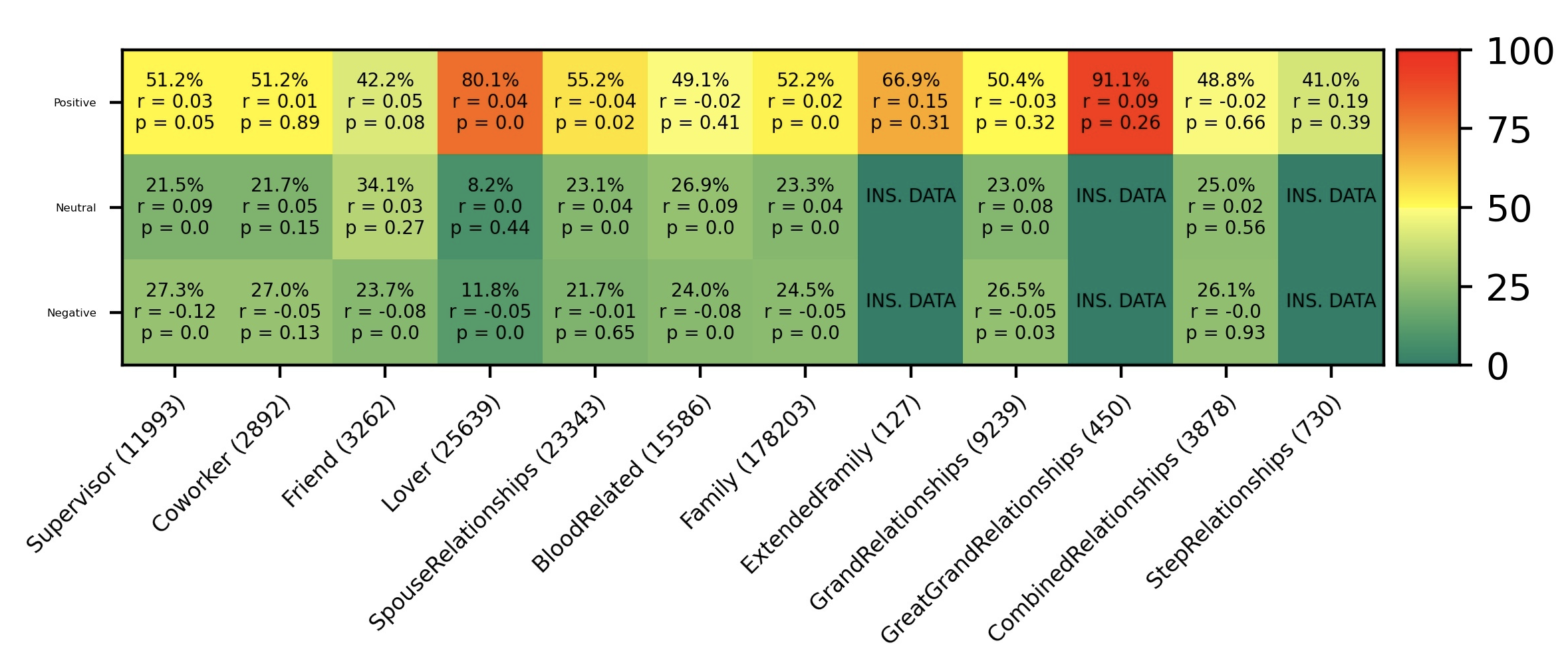}
\caption{Sentiment analysis of Indian tweets (top) and U.S. tweets (bottom) across different interpersonal categories using the culture-oblivious taxonomy. "INS. DATA" (Insufficient Data) indicates that at least one culture group did not contain enough tweets to allow statistical analysis.}
\label{fig:fine_grained_sent_combined}
\end{figure}

Fig.~\ref{fig:fine_grained_sent_combined} shows a breakdown of positive/negative sentiments expressed in tweets containing specific types of in-group relationships. Percentages indicate the proportion of tweets containing a given emotion, r values are the Rank Biserial Correlation for each emotion/relationship combination (with 0.0 indicating an identical distribution across both cultures), p-values show the level of statistical significance.

\section{FINANCIAL DISCLOSURES BASED ON RELATIONSHIP CATEGORY}\chngnum{19}

The results on comparison of financial disclosures across different interpersonal relationship categories are presented in Table~\ref{tab:cat_target_finance}. Due to the relative sparsity of financial-related disclosure data in the context of interpersonal relationships on Twitter, it is difficult to draw reliably conclusions from this data, and only some of the largest category groups are represented. However, interesting trends are visible in the frequency with which the word ``bought'' appears in the context of Family vs. Spousal vs. Lover relationships in the two cultures, with US tweeters referring most often to purchases in the context of spouse relationships while Indian tweeters use the word most in the context of lover relationships. Similarly interesting disclosure patterns exist for the words ``money'' and ``help''

\begin{table}[!h]
    \footnotesize
    \centering
    \begin{tabular}{p{3em} p{3em} p{3em} p{3em} p{3em} p{4em} p{4em} p{4em} p{4em}}
         & US-Family & Ind-Family & US-Grand & Ind-Grand & US-Spouse & Ind-Spouse & US-Lover & Ind-Lover \\
         & (1484) & (95) & (46) & (0) & (198) & (11) & (111) & (6) \\
        \toprule
        money & 14.69\% & 26.32\% & 10.87\% & 0.0\% & 7.58\% & 27.27\% & 9.90\% & 16.67\% \\ 
        buy & 18.46\% & 20.0\% & 8.70\% & 0.0\% & 17.68\% & 27.27\% & 27.03\% & 33.33\%\\ 
        help & 3.30\% & 14.74\% & 0.0\% & 0.0\% & 3.54\% & 0.0\% & 2.70\% & 16.67\%\\
        bought & 32.61\% & 28.42\% & 36.96\% & 0.0\% & 40.91\% & 9.09\% & 31.53\% & 50.0\%\\ 
        pay & 13.00\% & 7.34\% & 4.35\% & 0.0\% & 11.11\% & 0.0\% & 5.41\% & 0.0\% \\ 
        bank & 0.61\% & 2.11\% & 4.35\% & 0.0\% & 0.0\% & 9.09\% & 5.41\% & 0.0\% \\ 
        loan & 1.08\% & 0.0\% & 0.0\% & 0.0\% & 0.0\% & 0.0\% & 0.0\% & 0.0\% \\
        business & 0.88\% & 1.05\% & 0.0\% & 0.0\% & 1.01\% & 0.0\% & 0.90\% & 0.0\% \\ 
        card & 1.01\% & 1.05\% & 0.0\% & 0.0\% & 0.51\% & 0.0\% & 0.90\% & 0.0\% \\ 
        house & 4.18\% & 3.32\% & 0.0\% & 0.0\% & 1.52\% & 0.0\% & 1.80\% & 0.0\% \\ 
        give & 2.29\% & 1.05\% & 0.0\% & 0.0\% & 2.02\% & 0.0\% & 1.80\% & 16.67\%\\ 
        job & 1.15\% & 1.05\% & 0.0\% & 0.0\% & 2.53\% & 0.0\% & 0.90\% & 0.0\% \\ 
        salary & 0.0\% & 7.37\% & 0.0\% & 0.0\% & 0.0\% & 0.0\% & 0.0\% & 0.0\% \\ 
        car & 3.57\% & 0.0\% & 2.17\% & 0.0\% & 3.54\% & 18.18\% & 0.0\% & 0.0\% \\ 
        given & 0.0\% & 1.05\% & 0.0\% & 0.0\% & 0.0\% & 0.0\% & 0.0\% & 0.0\% \\ 
        tax & 1.28\% & 1.05\% & 2.17\% & 0.0\% & 0.51\% & 9.09\% & 0.0\% & 0.0\% \\ 
        afford & 1.89\% & 2.11\% & 0.0\% & 0.0\% & 3.54\% & 0.0\% & 0.0\% & 0.0\% \\ 
        insurance & 2.96\% & 0.0\% & 2.17\% & 0.0\% & 2.53\% & 0.0\% & 0.0\% & 0.0\% \\ 
        took & 1.35\% & 1.05\% & 0.0\% & 0.0\% & 1.52\% & 0.0\% & 1.80\% & 0.0\% \\
        payment & 0.27\% & 0.0\% & 2.17\% & 0.0\% & 0.0\% & 0.0\% & 0.0\% & 0.0\% \\ 
        income & 0.34\% & 0.0\% & 0.0\% & 0.0\% & 1.01\% & 9.0\% & 0.0\% & 0.0\% \\ 
        check & 0.40\% & 0.0\% & 0.0\% & 0.0\% & 0.51\% & 0.0\% & 0.0\% & 0.0\% \\ 
        \bottomrule
\end{tabular}
    \caption{\textbf{Financial disclosure based on relationship category.} We compare the frequency of each financially-related word across both cultural groups for three relationship categories.}
    \label{tab:cat_target_finance}

\end{table}

\section{USES OF THE TERM "SALARY" IN INDIA VS U.S. TWEETS}
\label{sec:appendix_salary}

Tweets from India frequently refer to salary in the context of providing money to one's parents. Tweets in the U.S. focus more on the tweeter's salary amount, and on desires to provide support to one's children. See Section \ref{sec:financial_quantitative} of the main paper. Table~\ref{tab:tweets_salary} exemplifies the  different use of the word salary in the context of financial disclosure. 

\begin{table}[!h]
    \centering
    \footnotesize
    \begin{tabular}{| p{20em} | p{20em} |}
        \hline
        \textbf{US} & \textbf{India}  \\
        \hline
        Something needs to change. I’m a teacher and pay \$18,600 a year for health care for my family of 3. This is more than 20\% of my salary pre taxes. @ewarren teachers need your help! \#fairteacherpay \#healthcare \#TeacherLife & Every time I get my salary I pay my credit card bills to my dad and it’s honestly an iconic moment because it truly reveals how much of a shopaholic I am \& my dad is just over it \\
        & \\
        Lmao my mama had money but she would not spend it on unnecessary shit for us . I could only do so much on my Wendy’s salary & Half salary of my mom and my dad (1 37 000 + 1 37 000.) \\
        & \\
        - my teacher salary was cut because “I was accidentally being overpaid” & i love spending my salary on my family, gosh it gives me some kind of warm feelings at heart. \\
        & \\
        I'm sure...my teacher salary wont let me hahaha & Where is my salary boss? Are u okay?\\
        & \\
         On a government employee salary, he provided college for all his kids! Pay everyone Bernie! Don’t discriminate!!! And cancel all mortgage and credit card debt too!!! & I gave my mom money from my salary every month. And she bought me new almari, and today new tilam. that money is for them not for me huhu apo la\\
        & \\
         & I spent 50\% of my salary to parent and family. Belanja makan. Check in. Shopping. Including my niece \& nephew things. Other stuffs.\\
        \hline
    \end{tabular}
    \caption{\textbf{Example tweet extracts demonstrating U.S. vs. India uses of "salary"} Samples were drawn from filtered tweets, meaning they referred specifically to financial disclosures.%
    }
    \label{tab:tweets_salary}
\end{table}

\section{WORD-LEVEL ANALYSIS OF TWEETS DISCLOSING LOCATIONS IN THE CONTEXT OF INTERPERSONAL RELATIONSHIPS}
Word-level analysis of location tweets (Table \ref{tab:location_analysis}) shows that U.S. tweeters are 3.7\% more likely to tweet about family relationships in conjunction with location disclosures and 5.4\% more likely to use the word love (which may be the result of Western tendencies to use "love" as a synonym for "like very much"). Indian tweeters were more likely to reference children and country in location-related tweets.

\begin{table}[h!]
    \centering
    \small
\begin{tabular}{p{7em} p{5em} p{5em}}
    & U.S. & India  \\
         & (6888) & (508)  \\
        \toprule
        \textbf{family} & \textbf{22.37\%} & \textbf{18.70\%} \\
        \textbf{love} & \textbf{13.24\%} & \textbf{7.87\%} \\
        \textbf{new} & \textbf{10.26\%} & \textbf{7.28\%} \\
        \textbf{children} & \textbf{4.03\%} & \textbf{6.50\%} \\
        friend & 5.28\% & 5.91\% \\
        happy & 4.45\% & 5.12\% \\
        \textbf{year} & \textbf{6.84\%} & \textbf{4.92\%} \\
        \textbf{christmas} & \textbf{9.55\%} & \textbf{4.53\%} \\
        \textbf{country} & \textbf{1.29\%} & \textbf{3.94\%} \\
        visit & 1.76\% & 2.76\% \\
        \bottomrule
\end{tabular}
\caption{\textbf{Location disclosure - word frequencies}. Percentages indicate number of tweets containing the specified word. Because of the structural differences between location and financial tweets, we were not able to apply the heuristic from Table \ref{tab:financial_disclosures} to create filtered versions of location-based tweets. Rows with particularly interesting discrepancies are bolded.}
\label{tab:location_analysis}
\end{table}

\section{Details of Auto-detected topics identified in Topical Disclosure Analysis}\label{sec:extra_topical_disclosure}

\begin{table}[!h]
\scriptsize
\begin{tabular}{| m{1.5cm}|m{4cm}|m{1.5cm}|m{4cm}|}
 \hline
 \textbf{U.S. Topic Name} & \textbf{Example Topic Words }  & \textbf{India Topic Name} & \textbf{Example Topic Words } \\
 \hline
Stories about Family &	brother, sister, girl, cousin, year, sweet, younger, sibling, twin, bore & Stories about Family	& feel, care, family, think, parent, sister, relationship, year, flight, girlfriend \\
\hline 
Complaining & talk, f**k, like, know, bitch, cause, wanna, listen, tell, n***a &  Celebrations &	friend, year, good, happy, best, dear, love, wish, family, love\\
\hline 
Gratitude & love, life, heart, thank, beautiful, amazing, world, know, forever, best & Expressing love & love, wait, friend, thank, life, drink, lover, water, familiy,like \\
\hline 
Celebrations & happy, birthday, year, wish, music, party, celebrate, surprise, cake, family & Patriotism & want, country, muslim, people, family, friend, children, india, know, need \\
\hline 
Christianity & feel, face, help, lose, sorry, hurt, felt, prayer, husband, better & Work & manager, mentor, coffee, mother, work, father, think, boss, time, today\\
\hline 
Politics & care, hope, kid, believe, children, future, people, need, help, change & schooling & teacher, help, school, hand, read, class, light, need, skin, leave\\
\hline 
Profane Narrative & miss, s**t, morn, tear, lma, hate, wake, laugh, like, dead & Reminiscing & friend, away, week, pass, laugh, today, beloved, family, memories, event \\
\hline 
Female Romantic partner & wife, marry, dream, catch, perfect, year, finish, single, hat, go & Complaining & hospital, brother, actress, shop, ticket, help, yoga, product, free, admit \\ 
\hline 
Schooling & school, supervisor, high, tomorrow, exciting, college, kid, work, today, go & other topics & - \\
\hline 
Social Media Activities & medium, nice, cute, free, store, invite, facebook, contact, request, post, channel & & \\
\hline 
Work & mentor, stuff, loud, happen, coffee, note, schedule, face,lead, newest & & \\
\hline 
Reminiscing & day, go, month, year, time, late, week, busy, visit, wrote & & \\
\hline 

\end{tabular}
\caption{LDA Topics identified in the U.S. and India along with 10 example topic words per topic. These topic names are manually coded by two researchers (one from the U.S. and another from India) collaboratively based on word clusters as well as sampled tweets associated with each topic.}
\label{table:us_india_topic_words_description}
\end{table}

We present example word per topics (i.e., word clusters) detected from interpersonal disclosures of U.S. and India in Table~\ref{table:us_india_topic_words_description}. \newtext{Our word clusters are representative of the topics and useful to possibly reproduce this work or re-detecting the same topics in future work. Each of these words appeared in multiple tweets and clustered by LDA.}\chngnum{21}

\end{document}